\documentclass[journal]{IEEEtran}
\ifCLASSINFOpdf
\else
\fi
\usepackage{amsmath}
\usepackage{amssymb}
\usepackage{amsthm}
\usepackage{bm}
\usepackage{mathrsfs}
\usepackage{graphicx}
\usepackage{epsfig}
\usepackage{caption}
\usepackage{subcaption}
\usepackage{enumerate}
\usepackage{cite}
\usepackage{setspace}
\usepackage{cancel} 
\usepackage{color}
\usepackage{wrapfig}
\usepackage{xspace}
\usepackage[]{hyperref}
\usepackage{times}
\usepackage{epstopdf}

\usepackage{array}
\usepackage{fixltx2e}
\usepackage{stfloats}

\usepackage{mydefs}
\usepackage{etoolbox}
\preto\subequations{\ifhmode\unskip\fi}

\theoremstyle{definition} 
\newtheorem{theorem}{Theorem}
\newtheorem{lemma}{Lemma} 
\newtheorem{proposition}{Proposition}
\newtheorem{corollary}{Corollary}
\newtheorem{definition}{Definition}

\begin{document}
%
\title{Higher-Order Nonlinear Complementary Filtering on Lie Groups}
%
%
%

\author{David~Evan~Zlotnik,~\IEEEmembership{Student Member,~IEEE,}
        James~Richard~Forbes,~\IEEEmembership{Member,~IEEE}%
\thanks{This research was supported by the National Science Foundation Award
Number 1550103.
}
\thanks{D.\ E.\ Zlotnik is with the Department
of Aerospace Engineering, University of Michigan, Ann Arbor,
MI, 48109 USA e-mail: \texttt{dzlotnik@umich.edu}.}
\thanks{J.\ R.\ Forbes is with the Department
of Mechanical Engineering, McGill University, Montreal,
QC, H3A 0C3, Canada e-mail: \texttt{james.richard.forbes@mcgill.ca}.}
}

\maketitle

\begin{abstract}
Nonlinear observer design for systems whose state space evolves on Lie groups is considered. The proposed method is similar to previously developed nonlinear observers in that it involves propagating the state estimate using a process model and corrects the propagated state estimate using an innovation term on the tangent space of the Lie group. In the proposed method, the innovation term is constructed by passing the gradient of an invariant cost function, resolved in a basis of the tangent space, through a linear time-invariant system. The introduction of the linear system completes the extension of linear complementary filters to nonlinear Lie group observers by allowing higher-order filtering. In practice, the proposed method allows for greater design freedom and, with the appropriate selection of the linear filter, the ability to filter bias and noise over specific bandwidths. A disturbance observer that accounts for constant and harmonic disturbances in group velocity measurements is also considered. Local asymptotic stability about the desired equilibrium point is demonstrated. A numerical example that demonstrates the desirable properties of the observer is presented in the context of pose estimation.
\end{abstract}

%
\IEEEpeerreviewmaketitle


\section{Introduction}

\IEEEPARstart{T}{he} kinematics and dynamics of many systems evolve on differential manifolds, rather than strictly in Euclidean space. Lie groups are a well known class of manifold that occur naturally in the study of rigid-body kinematics. Attitude kinematics, for example, evolve on the Lie group $SO(3)$, while pose kinematics evolve on the special Euclidean group $SE(3)$ \cite{book_Bullo_Lewis}. The development of observers for systems whose state evolves on a Lie group is therefore a highly practical exercise. A class of nonlinear Lie group observers has recently been developed \cite{paper_bonnabel_2008,paper_lageman_2008,paper_bonnabel_2009,paper_lageman_2010,paper_khosravian_2013,paper_khosravian_2015}.
The interest in these observers was, in part, sparked by the development of nonlinear observers for attitude estimation using first the unit quaternion \cite{paper_salcudean,book_nijmeijer,paper_thienel,paper_crassidis_2007} and later the rotation matrix element of $SO(3)$ directly \cite{paper_mahony_2005,paper_mahony_2008}.
Following \cite{paper_mahony_2008}, several nonlinear attitude observers that exploit the underlying $SO(3)$ Lie group structure have been developed \cite{paper_hua,paper_grip,paper_Khosravian_Namvar_2012}, as well as several $SE(3)$ based nonlinear observers \cite{paper_hua_2011,paper_hua_2015_gradient}. By working directly with the elements of $SO(3)$ and $SE(3)$, the attitude and pose are both globally and uniquely represented and thus issues associated with attitude parameterizations, such as non-uniqueness, are avoided \cite{chaturvedi2011rigid}. Nonlinear observers are attractive as they can often be shown to have strong stability properties \cite{paper_crassidis_2007,paper_lageman_2010,paper_khosravian_2013} and are, in general, computationally simpler than traditional estimation methods \cite{paper_lageman_2010}.

Lie group observer design methodology considered in the literature can often be described as possessing two distinct terms. The first is a copy of the nonlinear system while the second is an innovation term that serves to drive the estimated state towards the true state. This is analogous to the Luenberger observer for linear systems where the state estimate is propogated using a process model and a correction term alters the state estimate based on the error between the measured and estimated system output. However, as mentioned in \cite{paper_lageman_2010}, there is no canonical choice for the innovation term for Lie group observers, and as such, its selection must be carefully considered. A method to find suitable innovation terms is considered in \cite{paper_bonnabel_2008}, where symmetry preserving innovation terms are found via the moving frame method \cite{paper_bonnabel_2008,paper_khosravian_2013}. Alternatively, in \cite{paper_lageman_2008} and \cite{paper_lageman_2010} the innovation term is chosen based on the gradient descent direction of a selected invariant cost function and the observer is shown to be almost globally asymptotically stable about the point where the estimated state is equal to the true state. The work of \cite{paper_lageman_2010} was extended in \cite{paper_khosravian_2013} where a nonlinear Lie group observer for systems with homogeneous outputs and biased velocity measurements was proposed.


In this paper, a nonlinear observer whose system state evolves on a Lie group is considered. The approach taken is similar to previous Lie group observers presented in the literature, including \cite{paper_lageman_2010}, \cite{paper_khosravian_2013}, as the innovation is related to the gradient of a cost function. However, while the gradient of the cost function appears directly as the innovation term in \cite{paper_lageman_2010}, in this paper the innovation is based on the output of a linear time-invariant (LTI) system whose input is the gradient of a cost function resolved in a basis of the tangent space.
Analogous to the classical complementary filter, the proposed method can be understood as a nonlinear complementary filter on a Lie group.
Previous observers, including \cite{paper_lageman_2010} and \cite{paper_khosravian_2013}, are analogous to the classical complementary filter with first order sensitivity and complementary sensitivity transfer functions.
The introduction of the LTI system in the proposed observer allows for more general and complex higher-order filtering.
Consequently, to the authors' knowledge, the proposed method is the first nonlinear observer that fully extends the concept of linear complementary filtering to Lie group observer design.
This is highly practical, as it allows for the targeting of specific frequency bandwidths in the velocity and partial state measurements.
For example, applying a fast Fourier transform (FFT) to measurement data,
the frequency content of the measurement noise can be identified
and then mitigated by carefully constructing the LTI system.
Further, it is shown that by restricting the LTI sytem to the set of strictly positive real systems with feedthrough the strong stability properties of the nonlinear observer can be maintained. Passing nonlinear inputs through dynamic systems are considered in the context of rigid-body attitude estimation in \cite{paper_izadi_2014} and \cite{paper_izadi_2015}, and for rigid-body attitude control in \cite{paper_berkane_2017}.
In \cite{paper_berkane_2017} the nonlinear input is passed through a first-order dynamic system where the dynamic system is used to retain continuity in the case of discontinuities in the nonlinear input.
In \cite{paper_izadi_2014} and \cite{paper_izadi_2015}, a first-order dynamic system is employed and is derived via the Lagrange-D'Alembert principle and applied in continuous and discrete time.
The filters in \cite{paper_izadi_2014}, \cite{paper_izadi_2015}, and \cite{paper_berkane_2017} are first-order and consequently do not generalize to higher-order systems. Further, they were developed for the specific case of the Lie group $SO(3)$, while general Lie groups are considered in this paper.


The inclusion of an LTI system for filtering on the Lie group $SO(3)$ has been previously considered in \cite{zlotnik2016CDC}. This paper builds upon the results of \cite{zlotnik2016CDC} by considering a general Lie group as well as considering the effects of harmonic disturbances, both of which constitute significant contributions. Moreover, full state measurements of the rotation matrix element of $SO(3)$ is assumed in \cite{zlotnik2016CDC}, while the more realistic scenario of partial state measurements are assumed in this paper.
Further, the proposed method is a direct extension of the gradient observer proposed in \cite{paper_lageman_2010} in that the proposed observer reduces to the gradient observer with the selection of a particular static linear system. As in \cite{paper_khosravian_2013} and \cite{paper_khosravian_2015}, the case where velocity measurements are corrupted by constant bias is also considered. However, the solutions given in \cite{paper_khosravian_2013} and \cite{paper_khosravian_2015} are extended in this paper to include harmonic disturbances as well as constant bias. This is done by introducing a disturbance observer that incorporates an internal model of the harmonic disturbances. In practice, harmonic disturbances may be introduced due to mechanical vibration of a vehicle's structure or mechanical imbalance of rotors. For example, mounting an inertial measurement unit on a stiff but not infinitely rigid aerial vehicle frame would introduce harmonic disturbances into the angular velocity measurement \cite{pisu2008attitude}. The approach taken is similar to the method given in \cite{pisu2008attitude}, where an adaptive disturbance observer is proposed in the context of attitude estimation. In this paper a disturbance observer for use with any general Lie group is considered.

The main contribution of this paper is the design of the observer, which allows for greater design freedom compared to similar observers and can, with appropriate selection of the LTI system, result in enhanced rejection of measurement noise. It is shown that, provided the linear system is composed of a strictly positive real part and feedthrough, the proposed observer is locally asymptotically stable about a desired equilibrium point.

The remainder of this paper is as follows. Mathematical preliminaries are discussed in Sec.\ \ref{sec:preliminaries} including a discussion on Lie groups as well as Riemannian geometry. Observer design is considered in Sec.\ \ref{sec:filter}, where stability results are presented in Sec.\ \ref{sec:stability}. The disturbance observer is introduced in Sec.\ \ref{sec:disturbance}. The proposed observer is demonstrated in the context of rigid-body pose estimation in Sec.\ \ref{sec:example}. Finally, concluding remarks are given in Sec.\ \ref{sec:conclusion}.

\section{Mathematical Preliminaries and Notation} \label{sec:preliminaries}




We adopt notation from \cite{book_Bullo_Lewis}, \cite{book_bloch}, \cite{paper_lageman_2010} and \cite{paper_khosravian_2015}. Following \cite{paper_lageman_2010}, let $G$ and $\mathfrak{g}$ respectively denote a finite dimensional connected Lie group and its associated Lie algebra. An inner product on $\mathfrak{g}$ is denoted $\langle \cdot ,\cdot \rangle : \mathfrak{g} \times \mathfrak{g} \rightarrow \mathbb{R}$ with associated norm $|| \cdot ||_\mathfrak{g} = \sqrt{\langle \cdot, \cdot \rangle}$. Further, let $I \in G$ be the identity element and for $X,Y \in G$, and define the right and left translation maps by $R_X: G \rightarrow G, Y \mapsto YX$ and $L_X: G \rightarrow G, Y \mapsto XY$ \cite{book_Bullo_Lewis}. The tangent space of $G$ at any point $X \in G$ is denoted $T_XG$. Given $X \in G$ and $v \in \mathfrak{g}$, vectors in $T_XG$ may be expressed as $Xv$ or $vX$, where $Xv$ and $vX$ denote a simplified notation for $T_IL_Xv$ and $T_IR_Xv$, respectively \cite{paper_lageman_2010,book_bloch}. The adjoint map, denoted $\Ad_X : \mathfrak{g} \rightarrow \mathfrak{g}$, is a linear map where $\Ad_X(v) = T_X R_{X^{-1}} T_I L_X v$ for all $X \in G$ and $v \in \mathfrak{g}$ \cite{paper_khosravian_2015}.

Let the set $B = \{b_1,\ldots,b_n \}$, where $b_1,\ldots,b_n \in \mathfrak{g}$, be a basis of $\mathfrak{g}$. Then, for any $a \in \mathfrak{g}$, $a$ may be written as $a = S(\mbf{a})$ where $\mbf{a} = [\ a_1,\ldots,a_n \ ]^\trans \in \mathbb{R}^n$ and $S : \mathbb{R}^n \rightarrow \mathfrak{g}$ is such that $S(\mbf{a}) = \sum_{i=1}^n a_i b_i$.
A basis of $T_XG$ may be found by right translation of $B$ by $X$ such that $B_X = \{ b_1 X, \ldots, b_n X \}$ is a basis of $T_XG$. To simplify the results included in this paper the basis $B$ is assumed to be orthonormal, however, similar results follow for any arbitrary basis.


An inner product $\langle \cdot , \cdot \rangle_X : T_XG \rightarrow \mathbb{R}^+$ may be defined for the tangent space at each point $X$. When the inner product is smoothly varying the inner product is referred to as a Riemannian metric \cite{book_absil}. The inner product on $\mathfrak{g}$ determines a right invariant Riemannian metric by the following relationship
\beqarray
  \langle V(X) , U(X) \rangle_X & = &\langle  V(X)  X^{-1} ,  U(X) X^{-1} \rangle \nonumber \\
  & = & \langle v, u \rangle \nonumber \\
  & = & \mbf{v}^\trans \mbf{u} \label{eq:riemannian_metric}
\eeqarray
for all $X \in G$, and vector fields $V(X) = vX$ and $U(X) = uX$, where $v = S(\mbf{v}),u=S(\mbf{u}) \in \mathfrak{g}$ \cite{book_Bullo_Lewis}. The Riemannian metric of \eqref{eq:riemannian_metric} is right invariant in that $\langle V, U \rangle_X = \langle VY, UY \rangle_{XY}$ for all $X,Y \in G $ \cite{book_Bullo_Lewis}. Associated with the Riemannian metric is a unique torsion free and compatible affine connection $\nabla$, called the Levi-Civita connection, that assigns to each pair of vector field $V$ and $U$ a vector field $\nabla_V U$.

The gradient of a function $f: G \rightarrow \mathbb{R}$ is a vector field $\nabla f$ such that
\beq
  \mathscr{L}_V f(X) = \langle \nabla f(X), {V}(X)  \rangle_X, \ \forall {V}({X}) \in T_{{X}} G, \nonumber
\eeq
where $\mathscr{L}_V f(X)$ is the Lie derivative, or directional derivative, of $f$ along vector field $V$ at point $X$ \cite{book_absil,book_bloch,book_Bullo_Lewis}. The Riemannian Hessian, or Hessian operator, of $f$ at point $X$, is the symmetric linear mapping $H_{f(X)}: T_X G \rightarrow T_{X} G$ defined by \cite{book_absil}
\beq
  {H}_{f({X})}({V}({X})) = {\nabla}_{V} {\nabla}f({X}), \ \forall {V}({X}) \in T_{X} G. \label{eq:hessian}
\eeq

Consider now functions on $G \times G$. A function $f: G\times G \rightarrow \mathbb{R}$ is said to be symmetric if $f(X,Y) = f(Y,X)$ $\forall {X},{Y} \in G $ \cite{book_Bullo_Lewis}. The function $f$ is right invariant if $f({X}{Z}, {Y}{Z}) = f({X},{Y})$ $\forall {X},{Y},{Z} \in G$. Following \cite{book_Bullo_Lewis}, for a symmetric function $f: G \times G \rightarrow \mathbb{R}$ and ${Y} \in G$, define $f_{Y}: G \rightarrow \mathbb{R}$ as $f_{Y}({X}) = f({X},{Y})$. Then, the gradient of $f({X},{Y})$ with respect to ${X}$ and ${Y}$ are respectively defined as the unique vectors ${\nabla}_{{X}} f({X},{Y}) \in T_{X} G$ and ${\nabla}_{{Y}} f({X},{Y}) \in T_{Y} G$ such that
\beqarray
  \mathscr{L}_{V}f_{Y}({X}) & = & \langle {\nabla}_{{X}} f({X},{Y}), {V}  \rangle_{X}, \ \forall {V} \in T_{{X}} G \nonumber \\
  \mathscr{L}_{U}f_{X}({Y}) & = & \langle {\nabla}_{{Y}} f({X},{Y}), {U}  \rangle_{Y}, \ \forall {U} \in T_{{Y}} G. \nonumber
\eeqarray

\section{Nonlinear Complementary Filter} \label{sec:filter}

\subsection{Observer Design}

Consider the design of an observer for a system evolving on a Lie group. As before, let $G$ denote a Lie group with corresponding Lie algebra $\mathfrak{g}$, and let $X \in G$. The differential equations governing the trajectory $X(\cdot)$ is expressed as the left invariant system
\beq
  \dot{{X}}(t) = {X}(t) v(t), \label{eq:left_invariant}
\eeq
where $v(\cdot)$ is an exogeneous signal.
The quantity $v$ is often called the group velocity.
It is assumed that measurements of velocity $v(\cdot)$ and $\ell \in \mathbb{N}$, $l > 0$, partial measurements of the state $X$ are available as
\beqarray
  y_j(t) & = & h_j(N_j(t)X(t),\bar{{y}}_j), \ j \in \{ 1, \ldots, \ell\}, \\
  v_y(t) & = & v(t) + w(t), \label{eq:v_y}
\eeqarray
where $v_y \in \mathfrak{g}$ is the measurement of $v$, $w \in \mathfrak{g}$ is the noise associated with measurement $v_y$, and $N_j \in G$ is multiplicative noise associated with the measurement of $y_j$.
As in \cite{paper_lageman_2009}, the partial state measurements, or system outputs, $y_j$ are assumed to be elements of a manifold $\mathcal{M}$ and $\bar{y}_j \in \mathcal{M}$ are constant reference outputs. The mappings $h_j: G \times \mathcal{M} \rightarrow \mathcal{M}$ are assumed to be right actions of $G$ on $\mathcal{M}$ such that $h_j(I,y) = y$ and $h_j(X,h_j(Y,y)) = h_j(XY,y)$ for all $X,Y \in G$ and $y \in \mathcal{M}$ \cite{paper_lageman_2009,paper_khosravian_2015}. For simplicity of notation, define $\mathcal{M}^\ell = \mathcal{M} \times \cdots \times \mathcal{M}$ and let $y = ( y_1,\ldots, y_\ell ) \in  \mathcal{M}^\ell$, $\bar{y} = (\bar{y}_1,\ldots, \bar{y}_\ell) \in \mathcal{M}^\ell$, and $h : G \times \mathcal{M}^\ell \rightarrow \mathcal{M}^\ell$, $(X,y) \mapsto ( h_1(X,y_1),\ldots,h_\ell(X,y_\ell) )$.
%

To motivate the design of the observer presented in this paper, first consider the analogous system to \eqref{eq:left_invariant} on $\mathbb{R}$ given by
\beq
  \dot{x}(t) = v(t), \nonumber
\eeq
where $x \in \mathbb{R}$ and $v \in \mathbb{R}$ is some time-dependent exogenous signal. Suppose it is desired to build a filter to estimate the state $x$ from measurements of $x$ and $v$ given by
\beqarray
  y(t) & = & x(t) + n(t), \nonumber\\
  v_y(t) & = & v(t) + w(t), \nonumber
\eeqarray
where $n,w \in \mathbb{R}$ are, respectively, the noise associated with $y$ and $v_y$. The complementary filter is a simple method to fuse the measurements of $x$ and $v$, and is particularly effective when $y$ and $v_y$ have complementary noise characteristics \cite{paper_higgins_1975,paper_zimmermann_1992}.
Expressed in the frequency domain, the classical complementary filter is given by
\beqarray
  \hat{x}(s) & = & x(s) + \f{s}{s + H(s)} \f{w(s)}{s} + \f{H(s)}{s + H(s)} n(s) \nonumber \\
  & = & x(s) + S(s) \f{w(s)}{s} + T(s) n(s) \nonumber
\eeqarray
where $S(s)$ and $T(s)$ are the sensitivity and complementary sensitivity functions of the closed-loop system \cite{paper_higgins_1975,paper_zimmermann_1992,paper_mahony_2008}. The state-space representation of a complementary filter takes the form
\begin{subequations} \label{eq:complementary_filter}
\beqarray
  \dot{\hat{x}} & = & v_y - u \\
  \dot{\mbf{x}}_f & = & \mbf{A}_f \mbf{x}_f + \mbf{B}_f e \\
  u & = & \mbf{C}_f \mbf{x}_f + \mbf{D}_f e,
\eeqarray
\end{subequations}
where $(\mbf{A}_f,\mbf{B}_f,\mbf{C}_f,\mbf{D}_f)$ form a minimal state-space realization of $H(s)$, $\mbf{x}_f$ is the state associated with $H(s)$, $e = \hat{x} - y$ is the error between the state and output, and the temporal argument has been neglected for simplicity. Classical control methods can be used to design $H(s)$ such that $T(s)$ and $S(s)$ have desirable properties. When $n(s)$ is comprised of high frequency noise and $w(s)/s$ is comprised of low frequency noise, $H(s)$ is designed such that $T(s)$ and $S(s)$ are low-pass and high-pass filters, respectively. A simple method to accomplish this is to let $H(s) = k$, where $k\in(0,\infty)$. Then, $T(s)$ and $S(s)$ respectively become first order low and high-pass filters with cutoff frequencies of $k$ $(rad/s)$.

Motivated by \eqref{eq:complementary_filter}, the Lie group observer proposed in this paper takes the form
\begin{subequations} \label{eq:observer}
\beqarray
  \dot{\hat{X}} & = & \hat{X} v_y - u \hat{X}, \label{eq:X_hat} \\
  \dot{\mbf{x}}_f & = & \mbf{A}_f \mbf{x}_f + \mbf{B}_f \mbf{e} , \label{eq:x_f} \\
  \mbf{u} & = & \mbf{C}_f \mbf{x}_f + \mbf{D}_f \mbf{e},
\eeqarray
\end{subequations}
where $\hat{X}$ is the estimate of ${X}$, $ \mbf{H}(s) = \mbf{C}_f (s \mbf{1} - \mbf{A}_f)^{-1} \mbf{B}_f + \mbf{D}_f$ is a linear system with associated state $\mbf{x}_f \in \mathbb{R}^{n_f}$, $\mbf{e} \in \mathbb{R}^n$ is the input to $\mbf{H}(s)$, $u = S(\mbf{u}) \in \mathfrak{g}$, and $\mbf{u}$ is the output of the linear system.
%
The input $\mbf{e}$ is taken to be the representation of the gradient of a cost function resolved in basis $B_{\hat{X}}$. Let $g : \mathcal{M}^\ell \times \mathcal{M}^\ell \rightarrow \mathbb{R}^+$ denote a smooth symmetric cost function on $\mathcal{M}^\ell$ such that $g(h(\hat{X},\bar{y}),y)$ describes the error between predicted observations $h(\hat{X},\bar{y})$ and true observations $y = h(X,\bar{y})$. As in \cite{paper_lageman_2009}, it is assumed that $g$ is invariant under the right action $h(\cdot,\cdot)$ such that $g(h(X,a),h(X,b)) = g(a,b)$ for all $X\in G$ and $a,b \in \mathcal{M}^\ell$. A cost function on $G$ may be defined as $f : G \times G \rightarrow \mathbb{R}^+$, $f(\hat{X},X) = g( h(\hat{X},\bar{y}), h(X,\bar{y}) )$. As $g$ is invariant under the right action $h$, it follows that $f$ is a smooth symmetric right invariant function. Thus, $\mbf{e}$ is taken to be $ \mbf{e} = [ \nabla_{\hat{X}} f(\hat{X},X) ]_{B_{\hat{X}}}$, which is to say that $\mbf{e}$ is the representation of vector $ \nabla_{\hat{X}} f(\hat{X},X) \in T_{\hat{X}} G$ in the basis $B_{\hat{X}}$.

The proposed observer is composed of two coupled ordinary differential equations. The first, \eqref{eq:X_hat}, evolves directly on the underlying Lie group $G$, while the second, \eqref{eq:x_f}, is a linear system evolving on $\mathbb{R}^{n_f}$. Taken on its own, \eqref{eq:X_hat} shares the same structure as previous Lie group observers proposed in the literature, including \cite{paper_lageman_2010,paper_khosravian_2015}, in that it is composed of two terms, the first of which copies the nonlinear system dynamics of \eqref{eq:left_invariant} and the second is an innovation term that serves to drive the state estimate towards the true state. In fact, taking $\mbf{H}(s) = \mbf{1}$, the proposed observer reduces to
\beq
  \dot{\hat{X}} = \hat{X} v_y - {\nabla}_{\hat{X}} f(\hat{X},X), \label{eq:gradient_observer}
\eeq
the left gradient observer proposed in \cite{paper_lageman_2010}.

Noting the similarities in structure between the classical complementary filter \eqref{eq:complementary_filter} to the proposed observer \eqref{eq:observer}, the proposed method can be understood as a nonlinear complementary filter on the Lie group $G$. The similarities between Lie group observers of the form \eqref{eq:gradient_observer} and linear complementary filters with a constant transfer function $H(s) = k$ was first noted for the case of the Lie group $SO(3)$ in \cite{paper_mahony_2008}. The proposed observer, however, is analogous to a classical complementary filter on $\mathbb{R}$ for any general transfer function $H(s)$, rather than strictly for constant $H(s)$.
Therefore, the introduction of the linear system $\mbf{H}(s)$ in \eqref{eq:observer} completes the extension of linear complementary filters, with any general transfer function $H(s)$, to nonlinear complementary filters on Lie groups. In practice, $\mbf{H}(s)$ allows for greater freedom in the design of sensitivity and complementary sensitivity transfer functions when \eqref{eq:observer} is linearized. A constant transfer function only allows for simple first-order low- and high-pass filtering, while higher-order filtering can be accomplished with the appropriate selection of $\mbf{H}(s)$. This enhanced design freedom can be exploited to better reject measurement noise and improve performance of the nonlinear observer. It is shown in Sec.\ \ref{sec:stability} that the strong stability properties typical of nonlinear Lie group observers can be maintained even with the introuction of $\mbf{H}(s)$.

\subsection{Error Dynamics} \label{sec:error_dynamics}

As in \cite{paper_lageman_2010} and \cite{paper_khosravian_2015}, define the group error as $\tilde{X} = \hat{X} {X}^{-1}$ where $\tilde{X} = I$ when $\hat{X} = {X}$. As $f$ is right invariant, it follows that $f(\hat{X},{X}) = f_I(\tilde{X})$, where $f_I(\tilde{X}) \triangleq f(\tilde{X},I)$. To analyze the stability of the proposed observer it is helpful to determine the dynamics of $(\tilde{X},\mbf{x}_f)$. As is the case in \cite{paper_lageman_2010}, left invariant system dynamics along with the right invariance of the chosen Riemannian metric and cost function yield autonomous error dynamics. The autonomy of the error dynamics are established in the following proposition.
\begin{proposition}
  Consider trajectories of $(\hat{X},\mbf{x}_f)$ under \eqref{eq:observer}. Let $f: G\times G \rightarrow \mathbb{R}$ be a right invariant cost function and assume that $y$ and $\mbf{v}$ are measured exactly, that is, $y = h(X,\bar{y})$ and $\mbf{v}_y = \mbf{v}$.
  Then, dynamics associated with $(\tilde{X},\mbf{x}_f)$ are autonomous and are given by
  \begin{subequations} \label{eq:error}
  \beqarray
    \dot{\tilde{X}} & = & -u \tilde{X}, \\
    \dot{\mbf{x}}_f & = & \mbf{A}_f \mbf{x}_f + \mbf{B}_f  \mbf{e}, \\
    \mbf{u} & = & \mbf{C}_f \mbf{x}_f + \mbf{D}_f  \mbf{e},
  \eeqarray
  \end{subequations}
  where $\mbf{e} = \mbf{e}(\tilde{X}) = [ {\nabla}_{\tilde{X}} f_I(\tilde{X}) ]_{B_{\tilde{X}}}$.
\end{proposition}

\emph{Proof}
By Lemma 10 and Theorem 11 of \cite{paper_lageman_2010} the expression for $\dot{\tilde{X}}$ satisfies $\dot{\tilde{X}} = -T_{\hat{X}} R_{X^{-1}} u \hat{X} = -u \tilde{X}$.
%
To show that \eqref{eq:error} is autonomous, it is sufficient to show that $\mbf{e}$ depends only on $\tilde{X}$. Recall, $\mbf{e} = [ {\nabla}_{\hat{X}} f(\hat{X},{X}) ]_{B_{\hat{X}}}$ and therefore $ {\nabla}_{\hat{X}} f(\hat{X},{X}) = S(\mbf{e})\hat{X}$. As $f$ is right invariant and the gradient is defined with respect to a right invariant Riemannian metric, it follows that \cite[Lemma 16]{paper_lageman_2010}
\beqarraynn
	{\nabla}_{\tilde{X}} f_I(\tilde{X})  & = & T_{\hat{X}} R_{X^{-1}} {\nabla}_{\hat{X}} f(\hat{X},{X}) \\
  & = &  T_{\hat{X}} R_{X^{-1}} S(\mbf{e}) \hat{X} \\
  & = & S(\mbf{e}) \tilde{X}.
\eeqarraynn
Consequently, $ {\nabla}_{\tilde{X}} f_I(\tilde{X}) = S(\mbf{e}) \tilde{X}$ and thus $\mbf{e} = [ {\nabla}_{\tilde{X}} f_I(\tilde{X}) ]_{B_{\tilde{X}}} $. Therefore the components of $\mbf{e}$ depend only on $\tilde{X}$ and thus \eqref{eq:error} is autonomous. $\Box$

\subsection{Stability Results} \label{sec:stability}

In the stability results that follow restrictions will be made on the cost function, $f$, as well as the linear system $\mbf{H}(s)$. In particular, the cost function will be restricted to the set of right invariant error functions, as defined below, and the linear system is restricted to the set of strictly positive real systems with feedthrough.

\begin{definition}[Error function \cite{book_Bullo_Lewis}] \label{def:error_function}
 A smooth symmetric function $f : G \times G \rightarrow \mathbb{R}$ is an \emph{error function} about ${X} \in G$ if $f_{{X}}: G \rightarrow \mathbb{R}$ is smooth, proper, bounded from below, and $f_{X}$ satisfies
 \begin{enumerate}[(i)]
    \item $f_{X}({X}) = 0$,
    \item ${\nabla} f_{{X}}({X}) = 0$,
    \item ${H}_{f_{{X}}({X})}$ is positive definite.
 \end{enumerate}
\end{definition}
The properties of an error function are well established in \cite{book_Bullo_Lewis}, where the error function is labeled a ``tracking error function''.
A method for constructing error functions based on single variable cost functions on the output spaces is proposed in \cite{paper_khosravian_2013}. Another method for finding right invariant cost functions is discussed in \cite{paper_lageman_2010}.
\begin{definition}[Strictly Positive Real (SPR) Transfer Matrix \cite{paper_tao_1988}] \label{def:spr} A real, rational, strictly proper transfer matrix $\mbf{H}_{\text{spr}}(s)$ of the complex variable $s$ is SPR if
\begin{enumerate}
	\item $\mbf{H}_{\text{spr}}(s)$ is real for all real $s$ and all elements of $\mbf{H}_{\text{spr}}(s)$ are analytic in $Re\{ s\} \ge 0$,
	\item $\mbf{H}_{\text{spr}}(j\omega) + \mbf{H}_{\text{spr}}^\herm(j\omega) > \mbf{0}$ $\forall \omega \in (-\infty , \infty)$,
	\item $\lim_{\omega \rightarrow \infty} \omega^2 \{ \mbf{H}_{\text{spr}}(j\omega) + \mbf{H}_{\text{spr}}^\herm(j \omega) \} > \mbf{0}$.
\end{enumerate}
\end{definition}

Given Definition \ref{def:error_function} and Definition \ref{def:spr} it is now possible to present the main result of this section. As in \cite{paper_khosravian_2015} we require the existence of a faithful representation of the Lie group $G$ as a matrix Lie group.
\begin{theorem} \label{thm:local_asymptotic_stability}
  Consider trajectories of $(\tilde{X},\mbf{x}_f)$ under \eqref{eq:error}. Let $f$ be a right invariant error function about ${X}$, let $(\mbf{A}_f,\mbf{B}_f,\mbf{C}_f)$ be SPR where $\mbf{B}_f$ has full rank, and let $\mbf{D}_f \ge 0$. Assume that $y$ and $v$ are measured exactly, that is, $y = h(X,\bar{y})$ and $v_y = v$. Further assume that there exists a faithful representation of the Lie group $G$ as a matrix Lie group. Define $L = \sup\{ c \in \mathbb{R} \ | \ \tilde{X} \in \Omega_c \setminus \{I\} \implies \mbf{e} \neq \mbf{0} \}$, where $\Omega_c = \{ \tilde{X} \in G \ | \ f_I(\tilde{X}) \le c \}$. Then the following statements hold:
  \begin{enumerate}[(i)]
    \item \label{item:thm1_as} the equilibrium point $(\tilde{X},\mbf{x}_f) = (I,\mbf{0})$ is locally asymptotically stable;
    \item \label{item:thm1_ic} trajectories of $(\mbf{e},\mbf{x}_f)$ exponentially approach $(\mbf{0},\mbf{0})$ and $\tilde{X}$ asymptotically approaches $I$ for all initial conditions satisfying $V_1(\tilde{X}(0),\mbf{x}_f(0)) < L$, where $V_1$ is defined in \eqref{eq:V1}.
  \end{enumerate}
  %
\end{theorem}
\emph{Proof} See Appendix \ref{app:theorem1}.

The stability results presented in Theorem \ref{thm:local_asymptotic_stability} are valid for all $\mbf{H}(s) = \mbf{H}_\text{spr}(s) + \mbf{D}_f$, where $\mbf{H}_\text{spr}(s)$ is nonzero and $\mbf{D}_f \ge 0$. When $\mbf{H}_\text{spr}(s) = \mbf{0}$ a similar result can be found by further restricting $\mbf{D}_f$ to the set of positive definite marices, that is $\mbf{D}_f>0$. This can be accomplished by taking $f(\tilde{X},I)$ as the Lyapunov function and performing a similar analysis. This result is not presented here as it is equivalent to the stability results presented previously in \cite{paper_lageman_2010}.

The restriction on the set of initial conditions in item (\ref{item:thm1_ic}) can be interpreted as an estimate of the region of attraction of the equilibrium point $(\tilde{X},\mbf{x}_f) = (I,\mbf{0})$. Specifically, the estimate of the region of attraction is $\{ (\tilde{X},\mbf{x}_f) \in G \times \mathbb{R}^{n_f} \ | \ V_1(\tilde{X},\mbf{x}_f) \le c \}$, where $c < L$. It also follows, by Remark 6.13 of \cite{book_Bullo_Lewis}, that if $\tilde{X} = I$ is the only critical point of $f$, then the equilibrium point $(\tilde{X},\mbf{x}_f) = (I,\mbf{0})$ is globally asymptotically stable. Often it is the case that $f$ will have multiple critical points. In these instances, it is not possible to demonstrate global asymptotic stability. However, it may be possible to demonstrate almost global stability by placing further restrictions on the the error function as is done in \cite{paper_lageman_2010}. 

\section{Disturbance Observer} \label{sec:disturbance}

In the proofs in the previous section, it was assumed that the velocity term $v$ is measured exactly. However, in practice $v$ is often corrupted by noise and bias, as is the case for angular velocity measurements taken by inertial measurement units.
%
%
Suppose that the noise associated with $v_y$ in \eqref{eq:v_y}, $w = S(\mbf{w})$ where $\mbf{w} \in \mathbb{R}^n$, is composed of a linear combination of constant and harmonic signals. As such, $\mbf{w}$ may be written as the output of the linear system
\beq
  \dot{\mbf{x}}_d =  \mbf{A}_d \mbf{x}_d, \ \
  \mbf{w} =  \mbf{C}_d \mbf{x}_d, \label{eq:noise}
\eeq
where $\mbf{x}_d \in \mathbb{R}^{n_d}$ and $\mbf{A}_d$ is skew-symmetric.

Let $\mbfhat{x}_d$ denote the estimate of $\mbf{x}_d$ and consider the observer
\begin{subequations} \label{eq:observer_dist}
\beqarray
  \dot{\hat{X}} & = & \hat{X} v_y - \hat{X} \hat{w} - u \hat{X}, \label{eq:X_hat_disturbance}\\
  \dot{\mbf{x}}_f & = & \mbf{A}_f \mbf{x}_f + \mbf{B}_f  \mbf{e}, \\
  \mbf{u} & = & \mbf{C}_f \mbf{x}_f + \mbf{D}_f  \mbf{e},\\
  \dot{\mbfhat{x}}_d & = & \mbf{A}_d \mbfhat{x}_d + \rho \mbf{C}_d^\trans \mbfbar{e}, \label{eq:x_d} \\
  \mbfhat{w} & = & \mbf{C}_d \mbfhat{x}_d, \label{eq:w_hat}
\eeqarray
\end{subequations}
where $\rho > 0$, and $\hat{w} = S(\mbfhat{w}) \in \mathfrak{g}$. The input to \eqref{eq:x_d}, $\mbfbar{e} \in \mathbb{R}^n$, is such that $S(\mbfbar{e}) = \Ad_{\hat{X}}^*(S(\mbf{e}))$ where $\Ad_{\hat{X}}^*(\cdot)$ is the adjoint of the linear map $\Ad_{\hat{X}}(\cdot)$ such that $\langle u, \Ad_X(v) \rangle = \langle \Ad_X^*(u), v \rangle$ for all $v,u \in \mathfrak{g}$ and $X \in G$. The Lie group observer presented in \eqref{eq:observer_dist} is similar to the observer presented in \eqref{eq:observer}. However, the estimate of the disturbance, given by the disturbance observer in \eqref{eq:x_d} and \eqref{eq:w_hat}, is subtracted from the velocity measurement. Similar disturbance observers are used in \cite{pisu2008attitude} in the context of attitude estimation as well as in \cite{paper_Sanyal_Fosbury_Chaturvedi_Bernstein_2009} in the context of spacecraft attitude control.


The error dynamics associated with \eqref{eq:observer_dist} will be needed in the stability analysis that follows.

\begin{proposition}
  Define $\mbftilde{x}_d = \mbf{x}_d - \mbfhat{x}_d$ and define $\tilde{w} = S(\mbftilde{w})$ where $\mbftilde{w} = \mbf{w} - \mbfhat{w} \in \mathbb{R}^n$. Then, the dynamics associated with $(\tilde{X},\mbf{x}_f,\mbftilde{x}_d)$ are nonautonomous and are given by
  \begin{subequations} \label{eq:error_disturbance}
  \beqarray
    \dot{\tilde{X}} & = & \tilde{X}  \Ad_X(\tilde{w})  - u \tilde{X}, \label{eq:Xtilde_dot}\\
    \dot{\mbf{x}}_f & = & \mbf{A}_f \mbf{x}_f + \mbf{B}_f \mbf{e}, \\
  \mbf{u} & = & \mbf{C}_f \mbf{x}_f + \mbf{D}_f  \mbf{e},\\
  \dot{\mbftilde{x}}_d & = & \mbf{A}_d \mbftilde{x}_d - \rho \mbf{C}_d^\trans \mbfbar{e}, \\
  \mbftilde{w} & = & \mbf{C}_d \mbftilde{x}_d.
  \eeqarray
  \end{subequations}
\end{proposition}
\emph{Proof} By Lemma 10 and 11 of \cite{paper_lageman_2010}, the time derivative of $\tilde{X}$ satisfies $\dot{\tilde{X}} = T_{\hat{X}} R_{X^{-1}}( \hat{X} w -\hat{X} \hat{w} + u \hat{X})$. It follows then that
\beqarraynn
  \dot{\tilde{X}} & = & T_{\hat{X}} R_{X^{-1}}( \hat{X} \tilde{w} - u \hat{X}) \\
  & = & T_{\hat{X}} R_{X^{ -1}} T_I L_{\hat{X}} \tilde{w} - T_{\hat{X}} R_{X^{-1}} u \hat{X} \\
  & = & T_{X^{-1}} L_{\hat{X}} T_I R_{X^{-1}} \tilde{w} -  u \tilde{X} \\
  & = & T_I L_{\tilde{X}} T_{X^{-1}} L_X T_I R_{X^{-1}} \tilde{w} - u \tilde{X} \\
  & = & T_I L_{\tilde{X}} \Ad_X(\tilde{w}) - u \tilde{X} \\
  & = & \tilde{X} \Ad_X(\tilde{w}) - u \tilde{X}
\eeqarraynn
%
%
Taking the time derivative of $\mbftilde{x}_d$ gives
\beqarraynn
	\dot{\mbftilde{x}}_d & = & \dot{\mbf{x}}_d - \dot{\mbfhat{x}}_d \\
	& = & \mbf{A}_d \mbf{x}_d - \mbf{A}_d \mbfhat{x}_d - \rho \mbf{C}_d^\trans \mbfbar{e} \\
	& = & \mbf{A}_d \mbftilde{x}_d - \rho \mbf{C}_d^\trans  \mbfbar{e}.
\eeqarraynn
The error in the disturbance estimate is $\mbftilde{w} = \mbf{w} - \mbfhat{w} = \mbf{C}_d \mbf{x}_d - \mbf{C}_d \mbfhat{x}_d = \mbf{C}_d \mbftilde{x}_d$. The dynamics are nonautonomous due to the presence of the time dependent variable $X$ in the expression for $\dot{\tilde{X}}$.
$\Box$


The stability of the equilibrium point $(\tilde{X},\mbf{x}_f,\mbftilde{x}_d) = (I,\mbf{0},\mbf{0})$ is established in Theorem  \ref{thm:disturbance}. As before, it is required that $f$ be an error function and that $\mbf{H}(s) = \mbf{H}_\text{spr}(s) + \mbf{D}_f$, where $\mbf{H}_\text{spr}(s)$ is SPR and $\mbf{D}_f \ge 0$.

\begin{theorem} \label{thm:disturbance}
  Consider trajectories of $(\tilde{X},\mbf{x}_f,\mbftilde{x}_d)$ under \eqref{eq:error_disturbance}. Let $f$ be a right invariant error function about $X$ and let $(\mbf{A}_f,\mbf{B}_f,\mbf{C}_f)$ be SPR, where $\mbf{B}_f$ has full rank and $\mbf{D}_f \ge 0$.
  Assume that there exists a faithful representation of the Lie group $G$ as a Lie group, denoted $\mbs{\Phi}: G \rightarrow GL(m)$.
  Assume that $\mbs{\Phi}(X)$ and $v$ are bounded with respect to $||\cdot||_\frob$ and $ || \cdot ||_\mathfrak{g} $, respectively. Further assume that $\mbf{C}_d$ has full rank.
  Define $L = \sup\{ c \in \mathbb{R} \ | \ \tilde{X} \in \Omega_c \setminus \{I\} \implies \mbf{e} \neq \mbf{0} \}$, where $\Omega_c = \{ \tilde{X} \in G \ | \ f_I(\tilde{X}) \le c \}$. Then the following statements hold:
  \begin{enumerate}[(i)]
    \item \label{item:thm2_as} the equilibrium point $(\tilde{X},\mbf{x}_f,\mbftilde{x}_d) = (I,\mbf{0},\mbf{0})$ is locally uniformly asymptotically stable;
    \item \label{item:thm2_ic}trajectories of $(\tilde{X},\mbf{x}_f,\mbftilde{x}_d)$ converge asymptotically to $(I,\mbf{0},\mbf{0})$ for all initial conditions satisfying $V_3(\tilde{X}(0),\mbf{x}_f(0),\mbftilde{x}_d(0)) < L$, where $V_3$ is defined in \eqref{eq:V3}.
  \end{enumerate}
  %
  %
  %
\end{theorem}

\emph{Proof} See Appendix \ref{app:theorem3}.

As was the case for Theorem \ref{thm:local_asymptotic_stability}, the restriction on initial conditions in item (\ref{item:thm2_ic}) can be interpreted as an estimate of the region of attraction of the equilibrium point $(\mbftilde{X},\mbf{x}_f,\mbftilde{x}_d) = (I,\mbf{0},\mbf{0})$. This estimate is given by the set $\{ (\tilde{X},\mbf{x}_f,\mbftilde{x}_d) \in G \times \mathbb{R}^{n_f} \times \mathbb{R}^{n_d} \ | \ V_3(\tilde{X},\mbf{x}_f,\mbftilde{x}_d) \le c \}$, where $c < L$. It is also the case that if $L = +\infty$, then the equilibrium point is globally asymptotically stable.

\section{Pose Estimation Example} \label{sec:example}


In this section, the observer developed in this paper is applied to the problem of pose estimation and tested in simulation. Consider a rigid body rotating and translating in three dimensional space. The kinematics of the rigid body evolve on the Lie group
$SE(3)$
with corresponding Lie algebra $\mathfrak{se}(3)$ \cite{book_Bullo_Lewis}. The standard Euclidean matrix inner product, defined as $\trace{(\mbf{A}^\trans \mbf{B})}$ for all $\mbf{A},\mbf{B} \in \mathbb{R}^{n\times n}$, is taken as an inner product on $\mathfrak{se}(3)$. For convenience in notation elements in $SE(3)$ are identified by elements in $SO(3)\times \mathbb{R}^3$ by
\beq
  (\mbf{R},\mbf{r}) \mapsto \bma{c c} \mbf{R} & \mbf{r} \\ \mbf{0} & 1 \ema.
\eeq
%
Let $\mbf{T} = (\mbf{R},\mbf{r}) \in SE(3)$ denote the pose of the rigid body, where $\mbf{R} \in SO(3)$ is the attitude of the datum frame relative to the body frame, and $\mbf{r} \in \mathbb{R}^3$ is the position of the body relative to a datum resolved in the datum frame. In addition, let $\mbs{\omega} \in \mathbb{R}^3$ and $\mbf{v} \in \mathbb{R}^3$ denote the angular and translational velocities of the rigid body resolved in the body frame. Then, pose kinematics can be written as $\dot{\mbf{T}} = \mbf{T} \mbf{V}$, where $\mbf{V} = \bma{c c} \mbs{\omega}^\times & \mbf{v} \\ \mbf{0} & 0 \ema \in \mathfrak{se}(3)$, and $(\cdot)^\times: \mathbb{R}^3 \rightarrow \mathfrak{so}(3)$ such that
  \beq
   \mbs{\omega}^\times = \bma{c c c} 0 & -\omega_3 & \omega_2 \\ \omega_3 & 0 & -\omega_1 \\ -\omega_2 & \omega_1 & 0 \ema, \ \ \forall \mbs{\omega} = \bma{c} \omega_1 \\ \omega_2 \\ \omega_3 \ema \in \mathbb{R}^3. \nonumber
  \eeq
  %
  An orthonormal basis is chosen for $\mathfrak{g}$ and is given by $B = \{ \mbf{B}_1, \ldots, \mbf{B}_6 \}$. For this example, the basis is defined as
  \beqarraynn
    \mbf{B}_i & = & \bma{c c} \f{1}{\sqrt{2}} \mbf{e}_i^\times & \mbf{0} \\ \mbf{0} & 0 \ema, \ \forall i \in \{1,2,3\} \\
    \mbf{B}_{i} & = & \bma{c c} \mbf{0} & \mbf{e}_{i-3} \\ \mbf{0} & 0\ema, \ \forall i \in \{ 4,5,6\},
  \eeqarraynn
  where $\{\mbf{e}_1,\mbf{e}_2,\mbf{e}_3 \}$ is the standard basis of $\mathbb{R}^3$.

  \subsection{\texorpdfstring{$SE(3)$}{SO(3)} Observer Design}

  It is assumed that the rigid body is equipped with sensors that provide measurements of the group velocity, $\mbf{V}$, as well as the position of three reference points. The velocity is measured as
  \beq
  	\mbf{V}_y = \mbf{V} + \mbf{W}, \nonumber
  \eeq
  where $\mbf{W} = S(\mbf{w}) \in \mathfrak{se}(3)$ is the noise associated with measurement $\mbf{V}_y$. The reference vectors are measured as
  \beq
    \mbf{y}_j = \mbf{T}^{-1} \mbf{N}_j^{-1} \mbfbar{y}_j, \ j \in \{1,2,3 \},
  \eeq
  where $\mbf{y}_j \in \mathcal{M}$, $\mathcal{M} = \{ \mbf{x} = [\ x_1 \ x_2 \ x_3 \ x_4] \in \mathbb{R}^3 \ | \ x_4 = 1 \}$, is the partial state measurement, $\mbfbar{y}_j \in \mathcal{M}$ is a known reference, and $\mbf{N}_j = \exp(S(\mbf{n}_j))$, where $\mbf{n}_j \in \mathbb{R}^6$, is multiplicative noise associated with the vector measurement $\mbf{y}_j$.


  Let $\mbfhat{T}$ denote the estimate of $\mbf{T}$ and consider the function $f(\mbfhat{T},\mbf{T}) = \onehalf \sum_{j=1}^3 || \mbfhat{T}^{-1} \mbfbar{y}_j - \mbf{T}^{-1} \mbfbar{y}_j ||_2^2$. The function $f$ is a right invariant error function, satisfying all of the conditions in Definition \ref{def:error_function}, provided $\{ \mbfbar{y}_1,\mbfbar{y}_2,\mbfbar{y}_3 \}$ forms a basis of $\mathcal{M}$ \cite{paper_hua_2011}. The gradient of $f$ with respect to $\mbfhat{T}$ is
  $
    \mbs{\nabla}_{\mbfhat{T}} f(\mbfhat{T},\mbf{T}) = - \mbc{P}(\sum_{j=1}^3 \mbfhat{T}^{-\trans} ( \mbfhat{T}^{-1}  - \mbf{T})\mbfbar{y}_j \mbfbar{y}_j^\trans ) \mbfhat{T},
  $
  where $\mbc{P}(\cdot) : \mathbb{R}^{4\times4} \rightarrow \mathfrak{se}(3)$ is the orthogonal projection of $\mathbb{R}^{4 \times 4}$ onto $\mathfrak{se}(3)$ defined in \cite{paper_hua_2011}.
  %
  %
  %
  %
  The proposed observer, without the disturbance observer, given in \eqref{eq:observer} can be written as
  \begin{subequations} \label{eq:observer_example}
  \beqarray
      \dot{\mbfhat{T}} & = & \mbfhat{T} \mbf{V}_y - S(\mbf{u}) \mbfhat{T}, \label{eq:pose_observer} \\
      \dot{\mbf{x}}_f & = & \mbf{A}_f \mbf{x}_f + \mbf{B}_f \mbf{e}, \\
      \mbf{u} & = & \mbf{C}_f \mbf{x}_f + \mbf{D}_f \mbf{e},
  \eeqarray
  \end{subequations}
  where $\mbf{e}$ is the represenation of $- \mbc{P}(\sum_{j=1}^3 \mbfhat{T}^{-\trans} ( \mbfhat{T}^{-1}  - \mbf{T})\mbfbar{y}_j \mbfbar{y}_j^\trans )$ in the chosen basis $B$.



  To design $(\mbf{A}_f,\mbf{B}_f,\mbf{C}_f,\mbf{D}_f)$ it will be useful to examine the linearization of the error dynamics associated with \eqref{eq:observer_example}. These error dynamics are given by
  \begin{subequations} \label{eq:observer_example_error}
  \beqarray
      \dot{\mbftilde{T}} & = & \mbftilde{T} \mbf{T} S(\mbf{w}) \mbf{T}^{-1} - S(\mbf{u}) \mbftilde{R},  \\
      \dot{\mbf{x}}_f & = & \mbf{A}_f \mbf{x}_f + \mbf{B}_f  \mbf{e}, \\
     \mbf{u} & = & \mbf{C}_f \mbf{x}_f + \mbf{D}_f  \mbf{e}.
  \eeqarray
  \end{subequations}
%
 Let $\mbftilde{T} = \exp(S(\mbf{x}))$, where $\mbf{x} \in \mathbb{R}^6$. The  matrix $\mbftilde{T}$ can be perturbed about $\mbf{1}$ by letting $\mbf{x} = \mbfbar{x} + \delta \mbf{x}$, where $\mbfbar{x} = \mbf{0}$ is the nominal value of $\mbf{x}$. Writing $\mbftilde{T}$ as the exponential of $S(\delta \mbf{x})$ as a power series and neglecting higher order terms gives $\mbftilde{T} \approx \mbf{1} + S( \delta \mbf{x})$. Similarily, let $\mbf{w} = \mbfbar{w} + \delta \mbf{w}$ and $\mbf{n}_j = \mbfbar{n}_j + \delta \mbf{n}_j$, where $\mbfbar{w} = \mbfbar{n}_j = \mbf{0}$. It can be shown that the nonlinear input to the SPR filter in \eqref{eq:observer_example_error} can be expressed as $\mbf{e} \approx ( \mbf{M}_1 \delta \mbf{x} - \mbf{M}_2 \delta \mbf{n})$, where $\delta \mbf{n} = \bma{ c c c} \delta {\mbf{n}}_1^\trans & \delta {\mbf{n}}_2^\trans & \delta {\mbf{n}}_3^\trans \ema^\trans$, and $\mbf{M}_1 \in \mathbb{R}^{6 \times 6}$ and $\mbf{M}_2 \in \mathbb{R}^{6 \times 18}$ are full rank. 
%
Then \eqref{eq:observer_example_error} can be linearized as
\begin{subequations} \label{eq:lin}
\beqarray
	\delta \dot{\mbf{x}} & = & \delta \mbf{w}^\prime - \mbf{u} , \label{eq:lin_phi} \\
	\dot{\mbf{x}}_f & = & \mbf{A}_f \mbf{x}_f + \mbf{B}_f  ( \mbf{M}_1\delta \mbf{x} -  \mbf{M}_2 \delta \mbf{n} ), \\
	\mbf{u} & = & \mbf{C}_f \mbf{x}_f + \mbf{D}_f  ( \mbf{M}_1 \delta \mbf{x} -  \mbf{M}_2 \delta \mbf{n} ),
\eeqarray
\end{subequations}
where $ S(\delta \mbf{w}^\prime) = \Ad_{\mbf{T}} ( S(\delta \mbf{w}))$. To simplify the design of the filter, let $\mbf{H}(s) = \mbf{C}_f (s\mbf{1} - \mbf{A}_f)^{-1} \mbf{B}_f + \mbf{D}_f = H(s) \mbf{M}_1^{-1}$. Then, taking the Laplace transform of \eqref{eq:lin_phi} yields
\beqarraynn
	\delta \mbf{x}(s) & = & \f{s}{1 + H(s)} \f{\delta \mbf{w}^\prime(s)}{s} + \f{H(s)}{s + H(s)} \mbf{M}_1^{-1} \mbf{M}_2 \delta \mbf{n}(s) \nonumber \\
	& = & S(s) \f{\delta \mbf{w}^\prime(s)}{s} + T(s) \mbf{M}_1^{-1} \mbf{M}_2 \delta \mbf{n}(s). \label{eq:phi_tilde}
\eeqarraynn
A number of classical control techniques can now be used to design $H(s)$ such that $T(s)$ and $S(s)$ have desirable properties relative to the frequency content of $\delta \mbf{n}(s)$ and $\delta \mbf{w}^\prime(s)/s$. Different designs of $H(s)$ are explored in the following simulations.

\subsection{Simulations}

Let the angular and translational velocities of the rigid body be described by $\mbs{\omega} = -\pi^2 \cos(\pi/10 t)/60.0 \ [\ 1 \ 1 \ 1\ ]^\trans$ (rad/s) and $\mbf{v} = -0.1\pi^2 \cos(\pi/10 t)/60.0 \ [\ 1 \ 1 \ 1\ ]^\trans$ (m/s). The initial pose is set to
$\mbf{T}(0) = (\mbf{R}(0),\mbf{r}(0))$,
%
where $\mbf{R}(0) = \exp(\mbs{\phi}(0)^\times)$, $\mbs{\phi}(0) = [\ \pi/6 \ 0 \ 0 \ ]^\trans$, and $\mbf{r}(0) = [\ 1 \ 1 \ 1 ]^\trans$ (m). The three reference vectors are given by $\mbfbar{y}_1 = [ \ 1 \ 0 \ 0 \ 1 ]^\trans$, $\mbfbar{y}_2 = [ \ 0 \ 1 \ 0 \ 1 ]^\trans$, and $\mbfbar{y}_3 = [ \ 0 \ 0 \ 1 \ 1 ]^\trans$.
For the following simulations, the observer is initialized with $\mbfhat{T}(0) = \mbf{1}$ and $\mbf{x}_f(0) = \mbf{0}$. To highlight the design freedom that the linear filter affords, three different cases will be considered.

\subsubsection*{Case 1 (partial state measurement noise)}

The multiplicative noise associated with the reference vector measurements, $\mbf{N}_j = \exp(S(\mbf{n}_j))$, are constructed by selecting $\mbf{n}_j$ as
linear combinations of harmonic signals with frequencies in the range $[8\pi,16\pi]$ (rad/s) and amplitudes in the range $[0.05,0.4]$.
The undesirable effects of $\mbf{n}_j(t)$ can be mitigated by designing $T(s)$ as a low-pass filter with an appropriate cutoff frequency. Consider two versions of $H(s)$, $H_1(s) = k$ and $H_2(s) = b/(s+a)$, where $k,a,b \in (0,\infty)$. By selecting $H_1(s)$ the transfer function $T(s)$ is a first order low-pass filter of the form $T(s) = k/(s+k)$. This is equivalent to the pose observer proposd in \cite{paper_hua_2015_gradient}. Alternatively, selecting the SPR transfer function $H_2(s)$ results in a second order low-pass filter $T(s) = b/(s^2 + as + b)$. Selecting $k = 2$, $a = 6.2$, and $b = 9.7$ results in a cutoff frequency of $2$ (rad/s) for $T(s)$. Although $H_1(s)$ and $H_2(s)$ give the same cutoff frequency for $T(s)$, the second order low-pass filter rolls off at $-40$ dB per decade while the first order low-pass filter rolls off at $-20$ dB per decade. Consequently, it is expected that greater noise mitigation can be accomplished by selecting $H_2(s)$ over $H_1(s)$. Simulation results for the observer in \eqref{eq:observer_example} with both $H_1(s)$ and $H_2(s)$ are shown in Fig.\ \ref{fig:case1}. Referring to Fig.\ \ref{fig:case1}, the steady state error associated with the SPR filter $H_2(s)$ is indeed lower than that with a constant transfer function, as was expected. This indicates that superior noise mitigation is possible with the appropriate selection of an SPR filter.
%
%

\subsubsection*{Case 2 (input disturbance)} Suppose now that in addition to partial state measurement noise, the velocity measurement is corrupted by harmonic disturbances. Specifically, let $\mbf{w}(t) = [\ w_1(t) \  w_2(t) \ w_3(t) \ w_4(t) \ w_5(t) \ w_6(t)    \ ]^\trans$ be composed of harmonic signals such that
\beq
  w_i(t) = \alpha_i \sin(0.2\pi t) + \beta_i \cos(0.2\pi t), \ \forall i = 1,\ldots,6,
\eeq
where $\alpha_i,\beta_i \in [0.1,0.2]$ and $\alpha_i$ and $\beta_i$ have appropriate units. A simple method to mitigate the effect of $\mbf{w}$ is to add a notch filter at $0.2\pi$ (rad/s) to $S(s)$. In this way, the gain of $S(s)$ at $0.2\pi$ (rad/s) can be significantly reduced thereby attenuating $\mbf{w}$. Let $M(s) = (s^2 + 0.1 s + (0.2\pi)^2)/(s^2 + s + (0.2\pi)^2)$ be the notch filter, and consider a third version of $H(s)$, denoted $H_3(s)$. The filter $H_3(s)$ is designed by letting $S_3(s) = S_2(s) M(s)$, where $S_2(s) = s/(s+H_2(s))$, and solving for $H_3(s)$ by $S_3(s) = s/(s + H_3(s))$. Employing this method yields
\beqarraynn
      H_3(s) & = & \f{0.9s^3 + 15.25s^2 + 9.7 s + 3.8}{s^3 + 6.3 s^2 + 1.0 s + 2.45} \\
      & = & \f{9.6s^2 + 8.7s + 1.6}{s^3 + 6.315 s^2 + 1.0 s + 2.45} + 0.9 \\
      & = & H_{spr}(s) + D,
\eeqarraynn
where it can be shown that $H_{spr}(s)$ is SPR.
Simulation results for $H_1(s)$, $H_2(s)$, and $H_3(s)$ are shown in Fig.\ \ref{fig:case2}. The steady state performance of \eqref{eq:observer_example} with both $H_1(s)$ and $H_2(s)$ observed in case 1 has been lost. However, the results indicate that the inclusion of the notch filter in $H_3(s)$ is successful in mitigating the effects of the input disturbance. If in addition to harmonic disturbances a constant bias is added to $\mbf{w}$ such that
\beq
  w_i(t) = \alpha_i \sin(0.2\pi t) + \beta_i \cos(0.2\pi t) + b_i, \ \forall i = 1,\ldots,6,
\eeq
where $b_i \in \mathbb{R}$ is a constant, then the steady state performance of \eqref{eq:observer_example} is significantly degraded. This is shown in Fig.\ \ref{fig:case2_bias} where the previous simulation is repeated with $b_i \in [-0.5,0.5]$. The poor performance of \eqref{eq:observer_example} in this case motivates the introduction of the disturbance observer discussed in Sec.\ \ref{sec:disturbance}.

%
%

\subsubsection*{Case 3 (disturbance observer)}
\begin{figure}[!ht]
  \centering
  \includegraphics[width=0.5\textwidth]{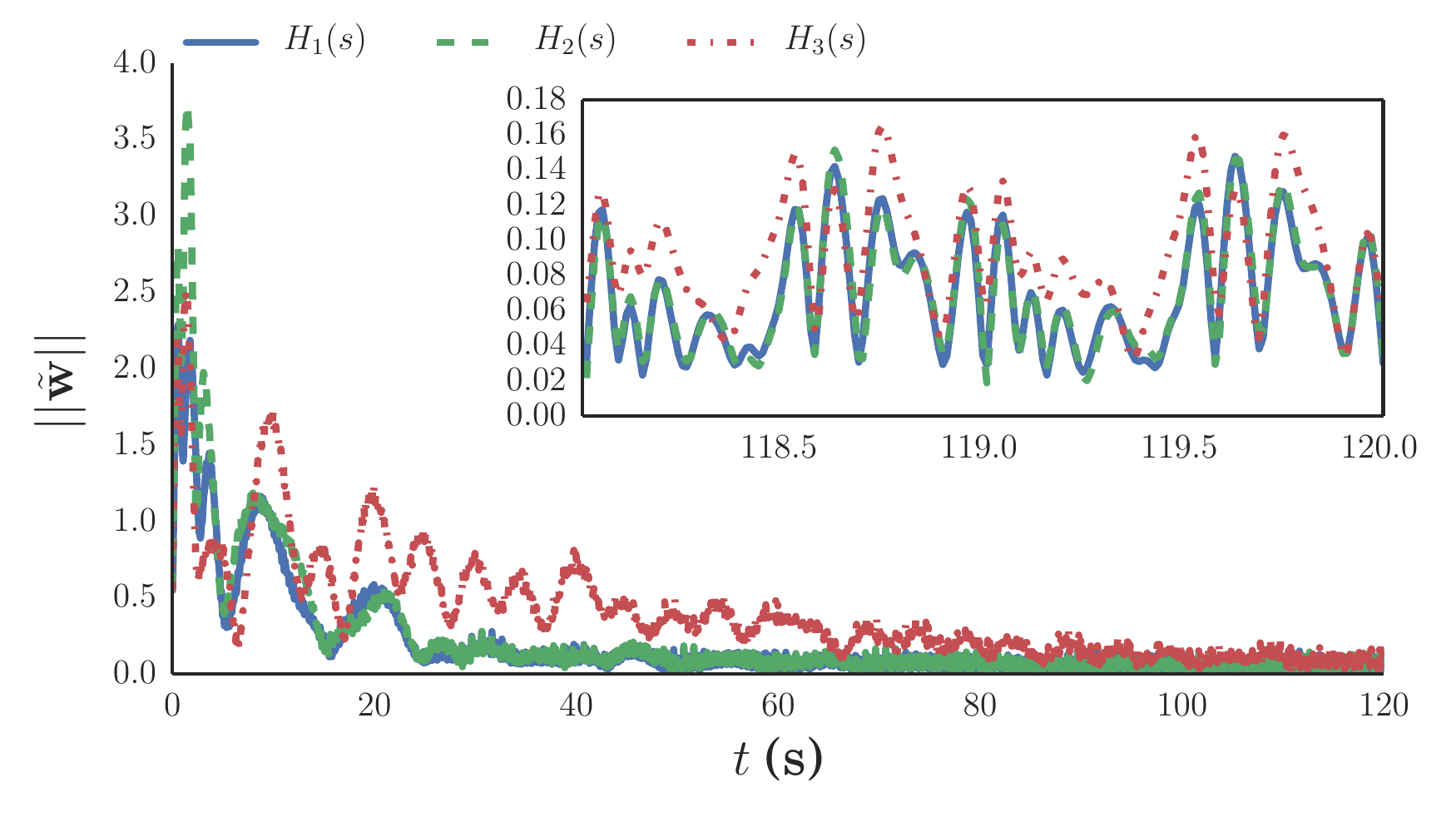}
  \caption{Error in the disturbance estimate approaches zero.}
  \label{fig:w_tilde}
\end{figure}

\begin{figure*}[!htbp]
  \begin{minipage}[b]{.5\linewidth}
    \centering
    \includegraphics[width=\textwidth]{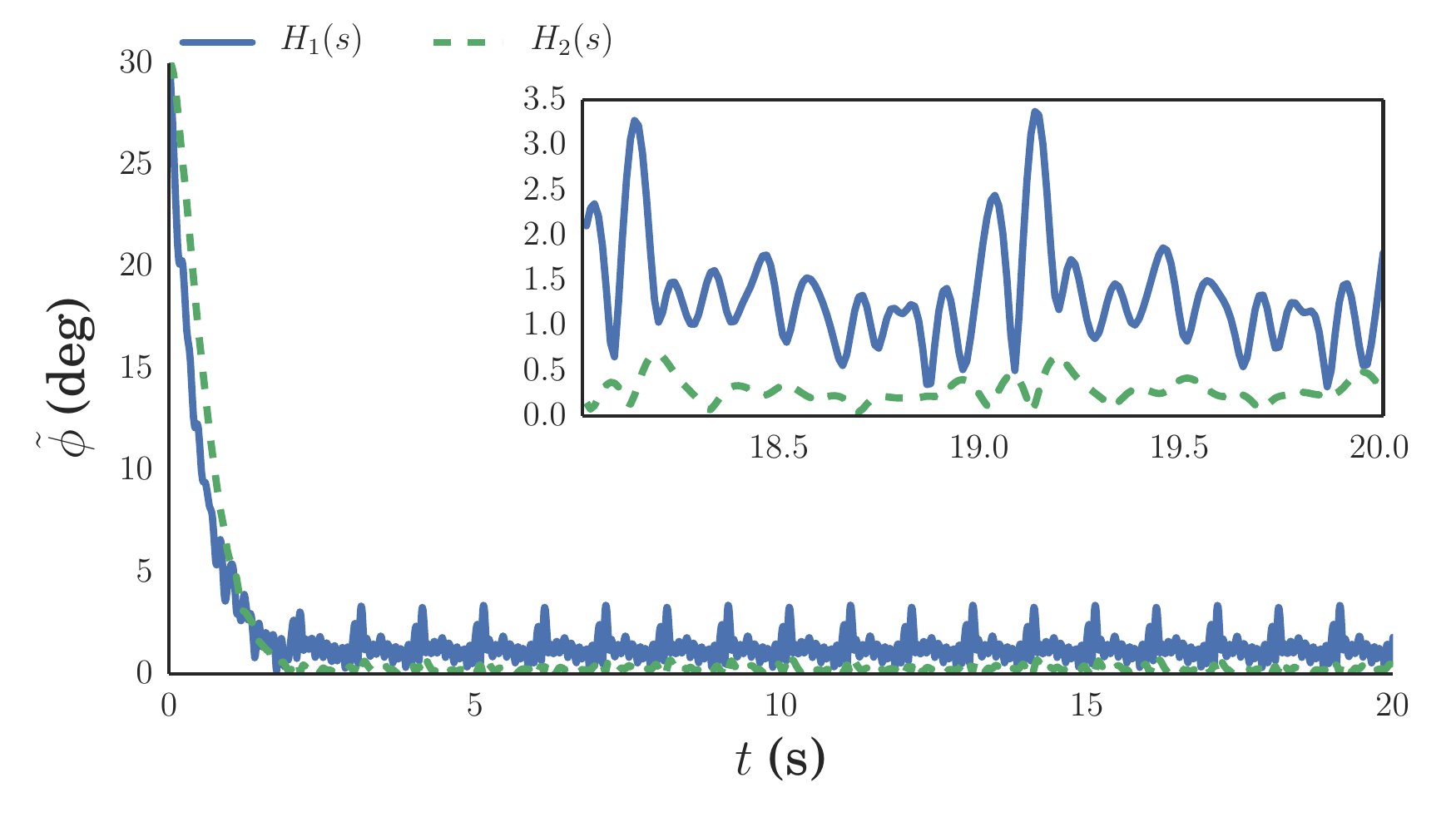}
  \end{minipage}%
  \begin{minipage}[b]{.5\linewidth}
    \centering
    \includegraphics[width=\textwidth]{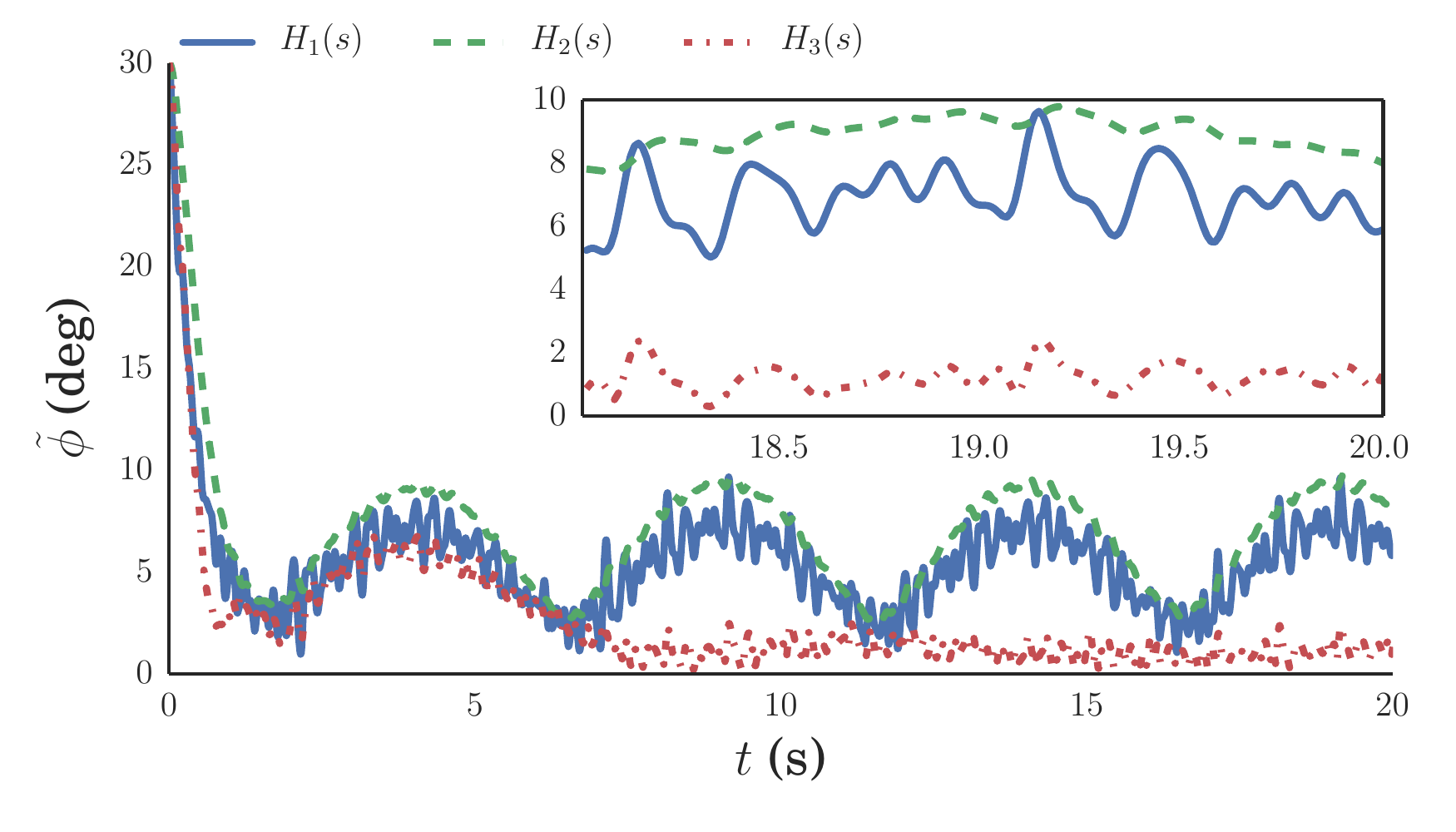}
  \end{minipage}\\
  \begin{subfigure}[b]{.5\linewidth}
    \centering
    \includegraphics[width=\textwidth]{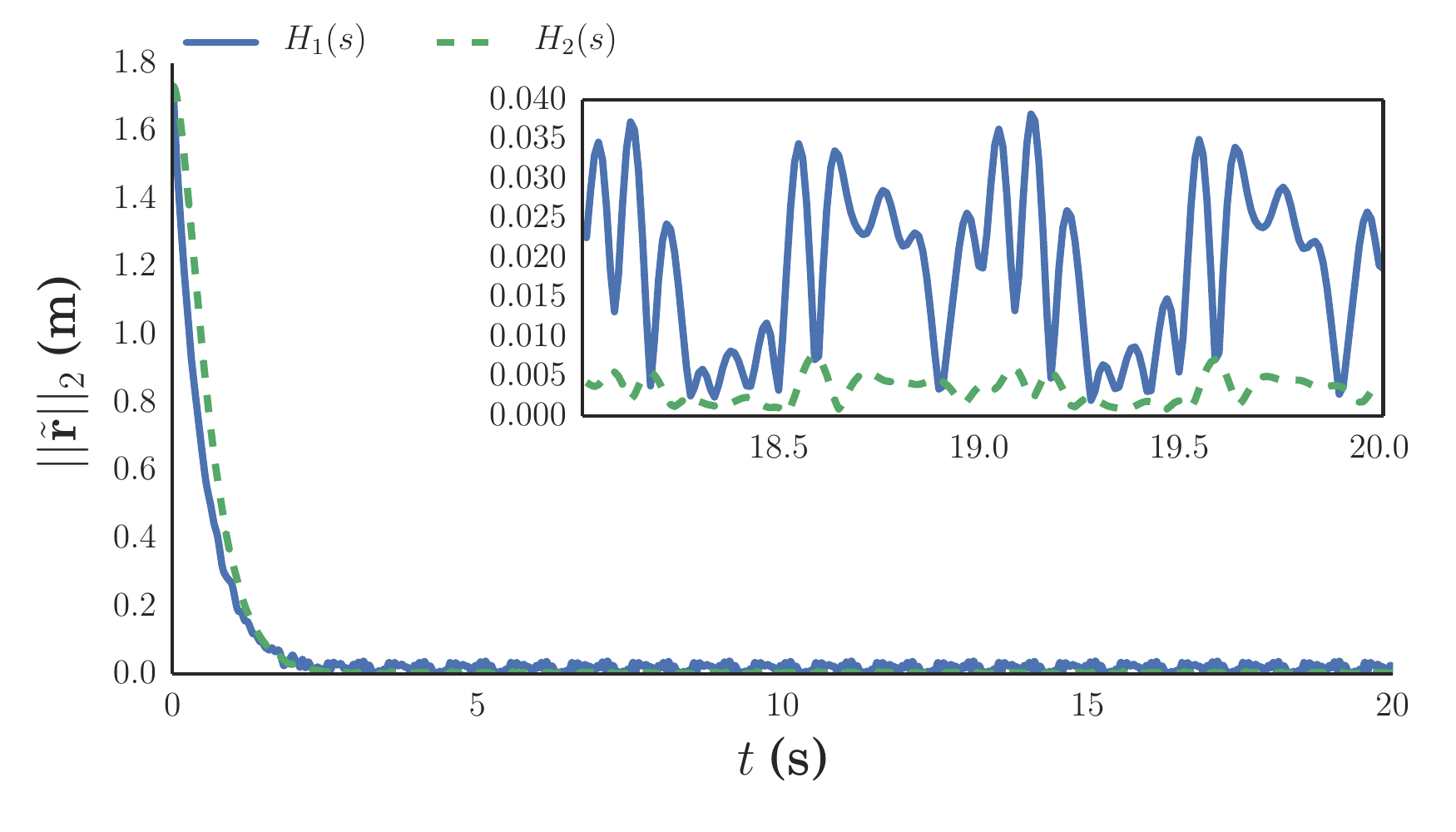}
    \caption{Simulation results for case 1.}
    \label{fig:case1}
  \end{subfigure}%
  \begin{subfigure}[b]{.5\linewidth}
    \centering
    \includegraphics[width=\textwidth]{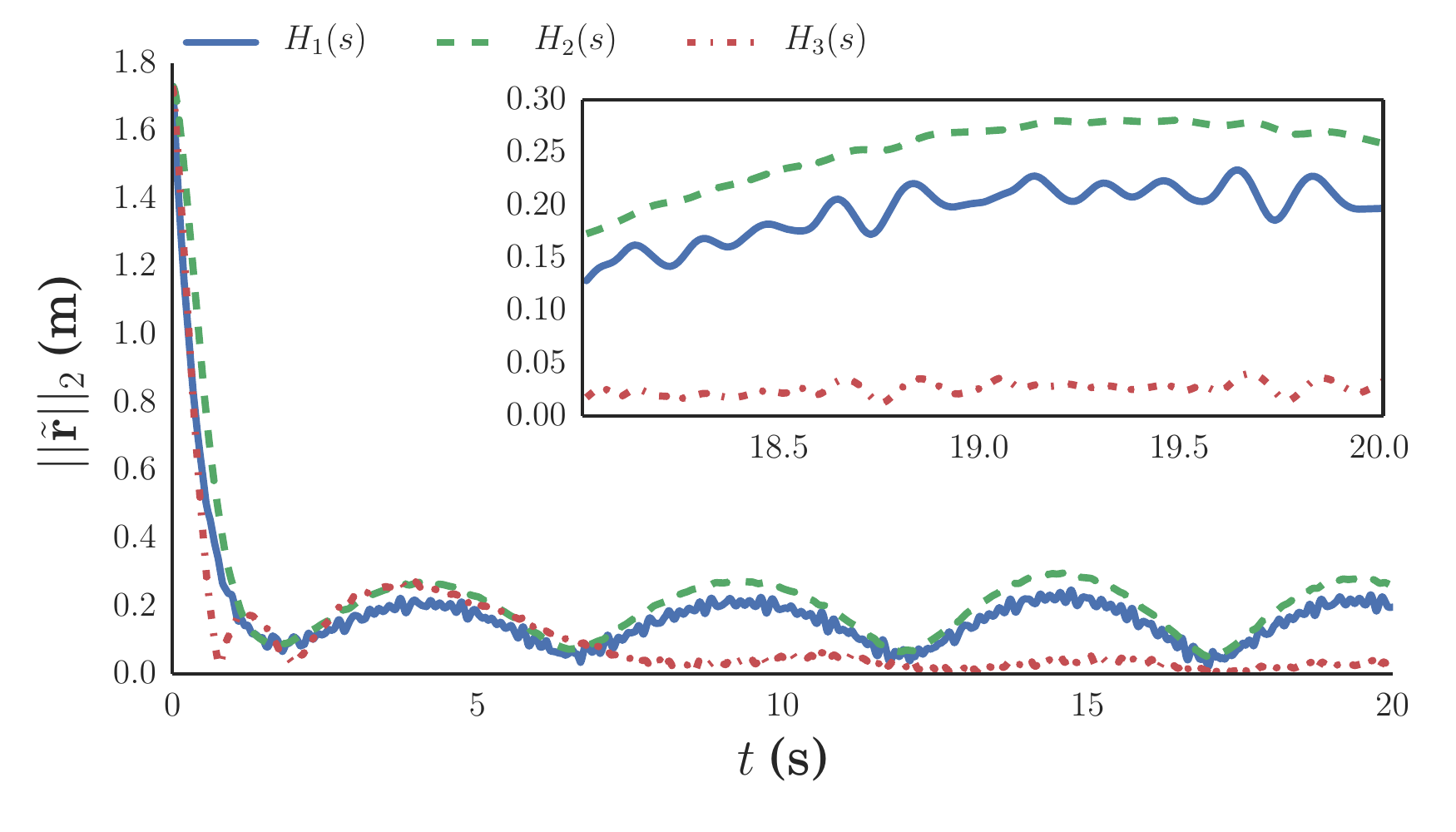}
    \caption{Simulation results for case 2.}
    \label{fig:case2}
  \end{subfigure} \\ [2em]
  \begin{minipage}[b]{.5\linewidth}
    \centering
    \includegraphics[width=\textwidth]{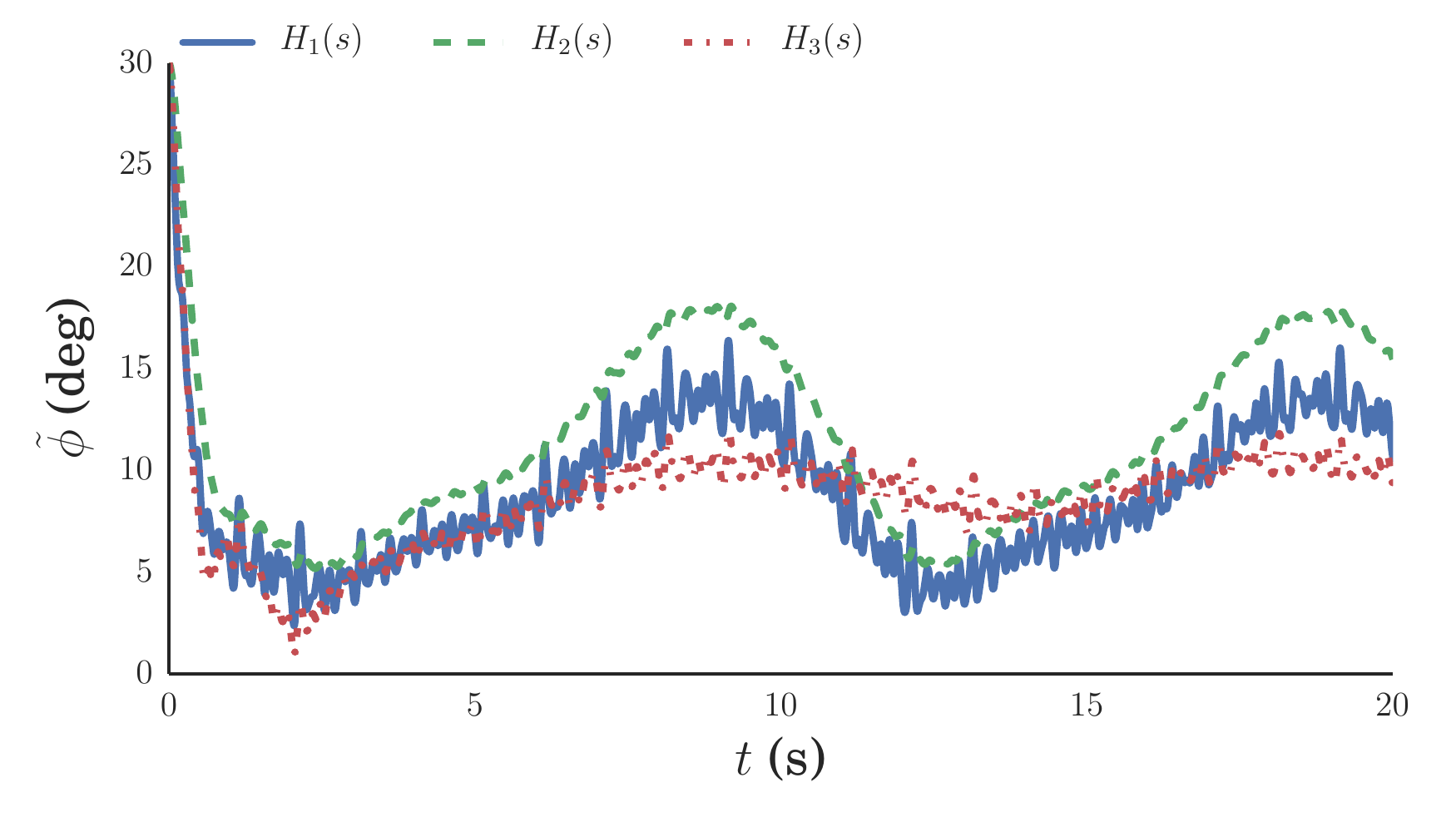}
  \end{minipage}%
  \begin{minipage}[b]{.5\linewidth}
    \centering
    \includegraphics[width=\textwidth]{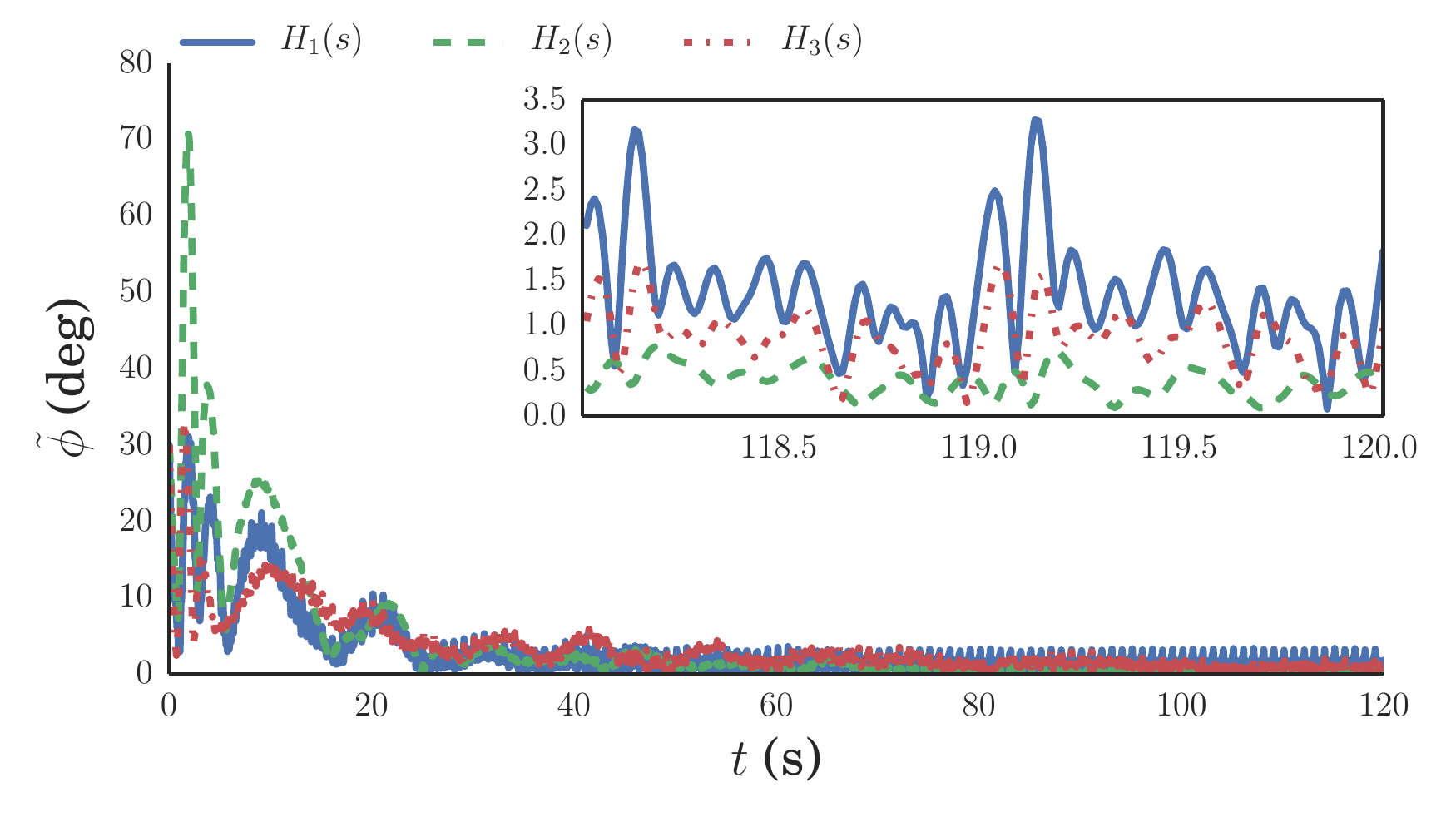}
  \end{minipage} \\
  \begin{subfigure}[b]{.5\linewidth}
    \centering
    \includegraphics[width=\textwidth]{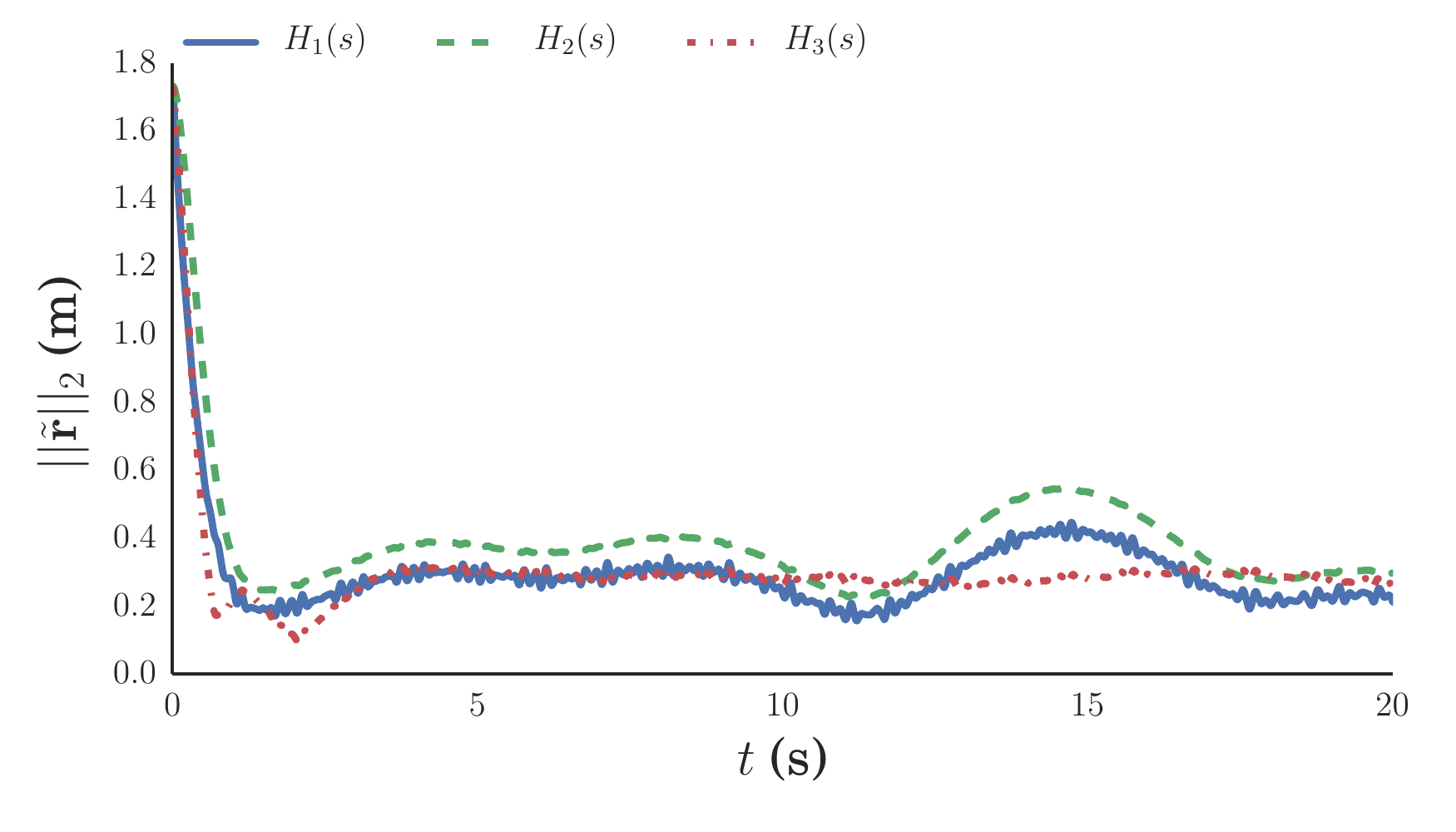}
    \caption{Simulation results for case 2 with constant bias.}
    \label{fig:case2_bias}
  \end{subfigure}
  \begin{subfigure}[b]{.5\linewidth}
    \centering
    \includegraphics[width=\textwidth]{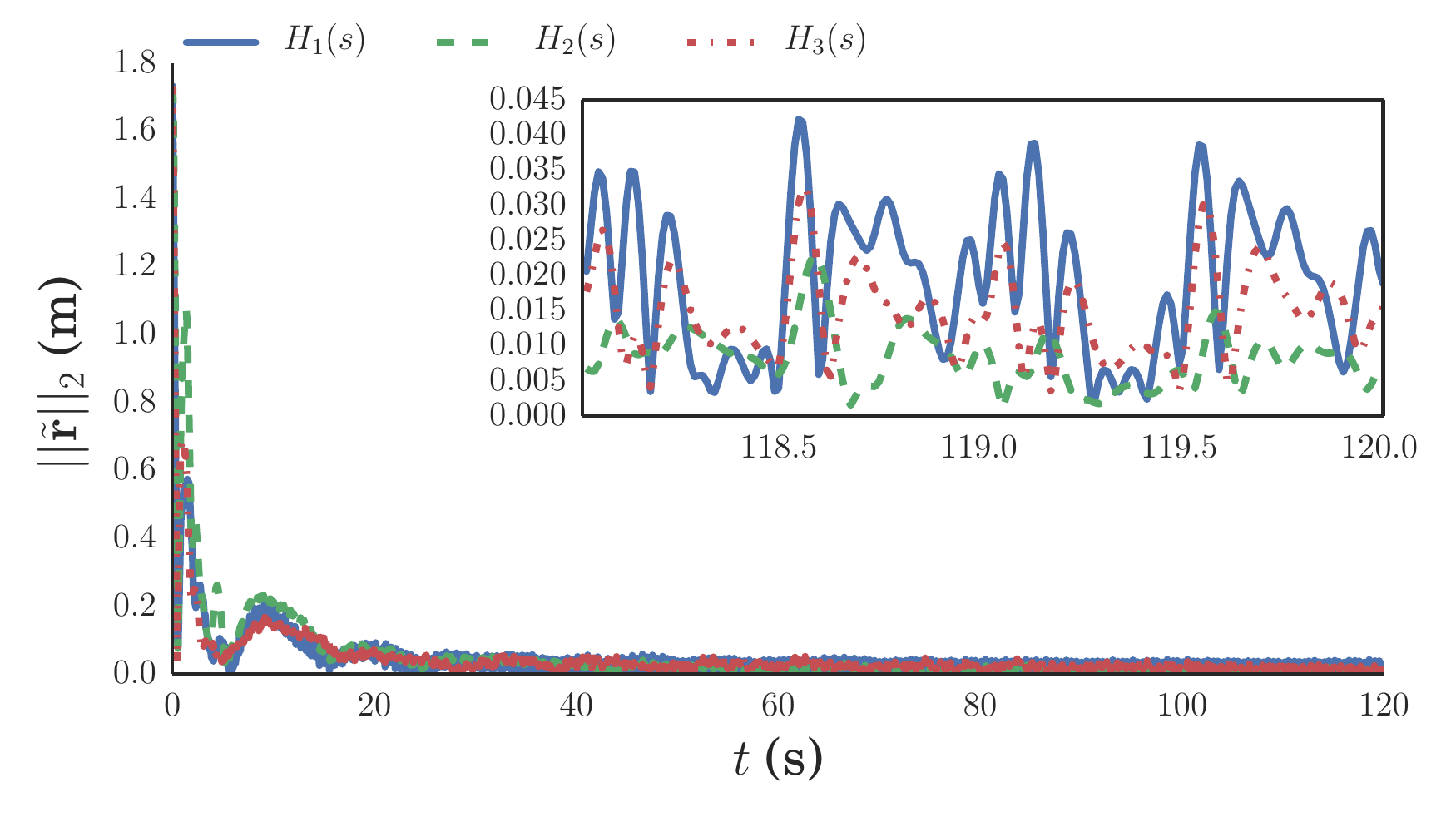}
    \caption{Simulation results for case 3.}
    \label{fig:case3}
  \end{subfigure}
  \caption{Time history of the error in the pose estimate for (a) case 1, (b) case 2, (c) case 2 with constant bias, and (d) case 3. The attitude error, $\tilde{\phi} = ||\mbstilde{\phi}||_2$, and position error, $||\mbftilde{r}||_2$, are extracted from $\mbftilde{T} =(\mbftilde{R},\mbftilde{r})$, where $\mbftilde{R} = \exp(\mbstilde{\phi}^\times)$. The steady-state error for the last two seconds is shown in the inset.}
  \label{fig:results}
\end{figure*}

In an effort to regain the steady-state performance of the proposed observer in the presence of velocity disturbances and constant bias, the disturbance observer presented in Sec.\ \ref{sec:disturbance} is now implemented. The disturbance, $\mbf{w}$, can be written as the output of the linear system in \eqref{eq:noise}, where $\mbf{A}_d = \text{diag}\{ \mbf{A}_1,\mbf{A}_2,\mbf{A}_3, \mbf{A}_4,\mbf{A}_6,\mbf{A}_6 \}$, $\mbf{C}_d = \text{diag} \{  \mbf{C}_1 , \mbf{C}_2 , \mbf{C}_3 ,  \mbf{C}_4, \mbf{C}_5, \mbf{C}_6\}$,
\beq
	\mbf{A}_i = \bma{c c c} 0 & 0 & 0 \\ 0 & 0 & 0.2 \pi \\ 0 & -0.2\pi & 0 \ema, \ \forall i=1,\ldots,6, \nonumber
\eeq
and $\mbf{C}_i = [ \ 1 \ 1/(0.2\pi) \ 0\ ]$ $\forall i = 1,\ldots,6$, with appropriate initial conditions. Then, the proposed observer on $SE(3)$ and the associated disturbance observer are given by
\begin{subequations}
\beqarraynn
	\dot{\mbfhat{T}} & = & \mbfhat{T}\mbf{V}_y - \mbfhat{T} S(\mbfhat{w}) - {S}(\mbf{u}) \mbfhat{T}, \\
	\dot{\mbf{x}}_f & = & \mbf{A}_f \mbf{x}_f + \mbf{B}_f  \mbf{e},\\
	\mbf{u} & = & \mbf{C}_f \mbf{x}_f + \mbf{D}_f  \mbf{e},\\
	\dot{\mbfhat{x}}_d & = & \mbf{A}_d \mbfhat{x}_d + \rho \mbf{C}_d^\trans \mbfbar{e}, \\
	\mbfhat{w} & = & \mbf{C}_d \mbfhat{x}_d,
\eeqarraynn
\end{subequations}
where $S(\mbfbar{e}) = \mbfhat{T} S(\mbf{e}) \mbfhat{T}^{-1}$. The results of a simulation with initial conditions $(\mbfhat{T},\mbf{x}_f,\mbfhat{x}_d) = ( \mbf{1},\mbf{0},\mbf{0} )$ and $\rho = 0.5$ is shown in Fig.\ \ref{fig:case3} and Fig.\ \ref{fig:w_tilde}. The pose estimation error is shown in Fig.\ \ref{fig:case3} while the error associated with the disturbance estimate is shown in Fig.\ \ref{fig:w_tilde}. Referring to Fig.\ \ref{fig:w_tilde}, it can be seen that the disturbance observer is successfull in tracking the true disturbance as $\mbftilde{w}$ approaches zero for all observers. Consequently, the steady-state attitude error observed in Fig.\ \ref{fig:case1} has been recovered, as shown in Fig.\ \ref{fig:case3}.

%

\section{Conclusion} \label{sec:conclusion}

Nonlinear observer design on Lie groups has been considered. This paper builds on previously developed nonlinear observers and the proposed method is a generalization of the gradient based Lie group observer proposed in \cite{paper_lageman_2010}. The observer has several desirable properties. First, like many recently developed nonlinear observers, the proposed method is provably locally asymptotically stable about the point at which the state estimate is equal to the true state. Secondly, the observer evolves directly on the underlying Lie group and thus captures the full nonlinear system dynamics. Third, a disturbance observer may be used to reject constant and harmonic disturbances in the velocity measurement. Finally, the introduction of an LTI system acting on the gradient of an invariant cost function allows for greater design freedom when compared to similar observers in the literature. Classical control techniques can be used to shape the sensitivity and complementary sensitivity transfer matrices of the linearized closed-loop system based on specific sensor noise characteristics. A numerical example demonstrated that the proposed observer performs admirably in the context of rigid-body pose estimation.


\appendices
\section{Proof of Theorem \ref{thm:local_asymptotic_stability}} \label{app:theorem1}

The proof of Theorem \ref{thm:local_asymptotic_stability} requires the following lemma and corollary.

\begin{lemma}[Kalman-Yakubovich-Popov (KYP) Lemma \cite{paper_wen}] \label{lemma:kyp}
Consider the LTI system
\beqarray
  \dot{\mbf{x}}_f & = & \mbf{A}_f \mbf{x}_f + \mbf{B}_f \mbf{u}_f, \nonumber \\
  \mbf{y}_f & = & \mbf{C}_f \mbf{x}_f, \nonumber
\eeqarray
where $\mbf{x}_f \in \mathbb{R}^{n_f}$, $\mbf{u}_f,\mbf{y}_f \in \mathbb{R}^{m_f}$, and $\mbf{A}_f$, $\mbf{B}_f$, and $\mbf{C}_f$ are appropriately dimensioned real matrices that form a minimal state-space realization. Moreover, assume that $\mbf{A}_f$ is Hurwitz. Then, the system is strictly positive real (SPR) if and only if there exists symmetric positive definite matrices $\mbf{P}_f,\mbf{Q}_f \in \mathbb{R}^{n_f \times n_f}$ such that
\begin{subequations} \label{eq:kyp_conditions}
\beqarray
  \mbf{P}_f \mbf{A}_f + \mbf{A}_f^\trans \mbf{P}_f & = & -\mbf{Q}_f \\
  \mbf{P}_f \mbf{B}_f & = & \mbf{C}_f^\trans. \label{eq:kyp_C}
\eeqarray
\end{subequations}
\end{lemma}

\begin{corollary} \label{cor:spr}
  Consider an LTI system with minimal state-space realization given by
  \begin{subequations} \label{eq:spr_system}
  \beqarray
    \dot{\mbf{x}}_f & = & \mbf{A}_f \mbf{x}_f + \mbf{B}_f \mbf{u}_f, \\
    \mbf{y}_f & = & \mbf{C}_f \mbf{x}_f + \mbf{D}_f \mbf{u}_f,
  \eeqarray
  \end{subequations}
  and define $\mathcal{L}(\mbf{x}_f) = \onehalf \mbf{x}_f^\trans \mbf{P}_f \mbf{x}_f$.
  If matrices $(\mbf{A}_f,\mbf{B}_f,\mbf{C}_f)$ satisfy the KYP equations \eqref{eq:kyp_conditions} then, regardless of the choice of $\mbf{D}_f$, the time derivative of $\mathcal{L}$ is given by
  %
  \beq
    \dot{\mathcal{L}}(\mbf{x}_f) = - \onehalf \mbf{x}_f^\trans \mbf{Q}_f \mbf{x}_f + \mbf{u}_f^\trans \mbf{y}_f - \mbf{u}_f^\trans \mbf{D}_f \mbf{u}_f. \nonumber
  \eeq
\end{corollary}

\emph{Proof} The proof follows directly from equations \eqref{eq:kyp_conditions} and \eqref{eq:spr_system}. $\Box$

Now a proof of Theorem \ref{thm:local_asymptotic_stability} is given, starting with item (\ref{item:thm1_as}) in Theorem \ref{thm:local_asymptotic_stability}. Consider the Lyapunov function candidate
\beq
  V_1(\tilde{X},\mbf{x}_f) = f(\tilde{X},I) + \mathcal{L}(\mbf{x}_f). \label{eq:V1}
\eeq
The derivative with respect to time of $V_1$ is
\beqarraynn
  \dot{V}_1(\tilde{X},\mbf{x}_f) & = & \langle {\nabla}_{\tilde{X}} f_I(\tilde{X}), \dot{\tilde{X}} \rangle_{\tilde{X}} + \dot{\mathcal{L}}(\mbf{x}_f) \\
  & = & -\langle S(\mbf{e}) \tilde{X},  S(\mbf{u})\tilde{X} \rangle_{\tilde{X}} + \dot{\mathcal{L}}(\mbf{x}_f)\\
  & = &  -\langle S(\mbf{e}), S(\mbf{u}) \rangle + \dot{\mathcal{L}}(\mbf{x}_f) \\
  & = &  -\mbf{e}^\trans  \mbf{u} + \dot{\mathcal{L}}(\mbf{x}_f).
\eeqarraynn
By application of Corollary \ref{cor:spr}, $\dot{V}_1$ is given by
\beq
  \dot{V}_1(\tilde{X},\mbf{x}_f) = -\mbf{e}^\trans \mbf{u} - \onehalf \mbf{x}_f^\trans \mbf{Q}_f \mbf{x}_f + \mbf{e}^\trans \mbf{u} - \mbf{e}^\trans \mbf{D}_f \mbf{e}. \nonumber
\eeq
Consequently, $\dot{V}_1(\tilde{X},\mbf{x}_f)  =  - \onehalf \mbf{x}_f^\trans \mbf{Q}_f \mbf{x}_f  - \mbf{e}^\trans \mbf{D}_f  \mbf{e}$.
%
By assumption $\mbf{D}_f \ge 0$, which implies $\dot{V}_1(\tilde{X},\mbf{x}_f) \le - \onehalf \mbf{x}_f^\trans \mbf{Q}_f \mbf{x}_f $ and thus $\dot{V}_1(\tilde{X},\mbf{x}_f) \le 0$ and $V_1(\tilde{X}(t),\mbf{x}_f(t)) \le V_1(\tilde{X}(0),\mbf{x}_f(0))$ for all $t \ge 0$. By Remark 11.11 of \cite{book_Bullo_Lewis}, $f$ is locally positive definite and $\tilde{X} = I$ is an isolated critical point of $f$.
By assumption, there exists a faithful representation of $G$ as a matrix Lie group. This implies that there exists $m>0$ and a mapping $\mbs{\Phi} : G \rightarrow GL(m)$ such that $\mbs{\Phi}(G)$ is a matrix Lie group \cite{paper_khosravian_2015}.
Following the proof of Theorem 5.1 in \cite{paper_khosravian_2015}, this implies that there exists a set $B_r = \{ \tilde{X} \in G \ | \ d(\tilde{X}) \le r \}$ about $\tilde{X} = I$, where $d(\tilde{X}) = || \mbf{1} - \mbs{\Phi}(\tilde{X})||_\frob$, such that for all $\tilde{X} \in B_r$, $f(\tilde{X},I)$ is positive definite and $\tilde{X}= I$ is the only critical point of $f$ in $B_r$. Further, this implies that $V_1(\tilde{X},\mbf{x}_f)$ is positive definite in the set $\bar{B}_r = \{ (\tilde{X},\mbf{x}_f) \in G \times \mathbb{R}^{n_f} \ | \ \ell(\tilde{X},\mbf{x}_f) \le r \}$, where $\ell(\tilde{X},\mbf{x}_f) = d(\tilde{X}) + ||\mbf{x}_f||_2$, and $\tilde{X} = I$ is the only critical point of $f$ in $\bar{B}_r$.


A corollary to LaSalle's invariant set theorem will now be used to prove local asymptotic stability \cite[p.\ 128]{khalil2002nonlinear}. Let $\mathcal{S} = \{ (\tilde{X},\mbf{x}_f) \in \bar{B}_r \ | \ \dot{V}_1(\tilde{X},\mbf{x}_f) = 0 \}$. It will now be shown that the only solution that can stay identically in $\mathcal{S}$ is the solution $(\tilde{X},\mbf{x}_f) = (I,\mbf{0})$. For all $(\tilde{X},\mbf{x}_f) \in \mathcal{S}$, $\mbf{x}_f \equiv \mbf{0}$. With $\mbf{x}_f \equiv \mbf{0}$, it follows that $\dot{\mbf{x}}_f \equiv \mbf{0}$. This implies that $\mbf{B}_f  \mbf{e} \equiv \mbf{0}$. Since $\mbf{B}_f$ has full rank by assumption, $\mbf{B}_f  \mbf{e} \equiv \mbf{0}$ implies that $\mbf{e} \equiv \mbf{0}$ and consequently ${\nabla}_{\tilde{X}} f_I(\tilde{X}) \equiv 0$. As the only critical point of $f$ in $\bar{B}_r$ is the point $\tilde{X} = I$, it follows that ${\nabla}_{\tilde{X}} f_I(\tilde{X}) \equiv 0$ implies $\tilde{X} = I$. Thus, by Corollary 4.1 of \cite[p.\ 128]{khalil2002nonlinear} the equilibrium point $(\tilde{X},\mbf{x}_f) = (I,\mbf{0})$ is locally asymptotically stable. This proves item (\ref{item:thm1_as}).


The following corollary will be required for the proof of item (\ref{item:thm1_ic}).

\begin{corollary} \label{cor:u_f}
  Consider the gradient vector field ${\nabla}f_I(\tilde{X}) = S(\mbf{e}) \tilde{X}$ and let $\dot{\tilde{X}} = S(\mbf{q}) \tilde{X}$, where $\mbf{q} \in \mathbb{R}^n$. Then, the derivative with respect to time of $\mbf{e}$ is given by
  \beq
    \dot{\mbf{e}} = \mbf{H}(\tilde{X}) \mbf{q} - \mbs{\xi}, \nonumber
  \eeq
  where $\mbf{H}(\tilde{X})$ is the matrix representation in basis $B_{\tilde{X}}$ of the Riemannian Hessian operator $H_{f_I(\tilde{X})}(\cdot)$ at point $\tilde{X}$, and $\mbs{\xi}\in\mathbb{R}^n$. Moreover, there exists a finite constant $m < \infty$ such that $\mbs{\xi}$ satisfies $ || \mbs{\xi} ||_2 \le m || \mbf{q} ||_2 || \mbf{e} ||_2$.
\end{corollary}

\emph{Proof} Let $\tilde{X}$ be a trajectory under the ordinary differential equation $\dot{\tilde{X}} = S(\mbf{q}) \tilde{X}$, and consider the gradient vector field ${\nabla} f_I(\tilde{X})$. Let $\Gamma$ be a vector field along the curve $\tilde{X}$ such that $\Gamma(t) = {\nabla} f_I(\tilde{X}(t))$. The covariant derivative of $\Gamma$ along $\tilde{X}$ is given by \cite[p. 139]{book_Bullo_Lewis},
\beq
  \f{\mathrm{D} \Gamma }{\mathrm{d}t} = S(\dot{\mbf{e}}) \tilde{X} + Q( S(\mbf{q}) , S(\mbf{e} ) ) \tilde{X}, \label{eq:f_cov}
\eeq
where $Q: \mathfrak{g} \times \mathfrak{g} \rightarrow \mathfrak{g}$ is the unique bilinear mapping associated with the Levi-Civita connection such that for any two right invariant vector fields $V(X) = vX$ and $U(X) = uX$, $\nabla_V U = Q(v,u)X$.
By definition the covariant derivative satisfies
\beq
  \f{\mathrm{D} \Gamma }{\mathrm{d}t} = {\nabla}_{\dot{\tilde{X}}} {\nabla}f_I(\tilde{X}), \nonumber
\eeq
and from the definition of the Hessian operator in \eqref{eq:hessian} it follows that
\beq
  \f{\mathrm{D} \Gamma }{\mathrm{d}t} =H_{ f_I(\tilde{X}) } (  \dot{\tilde{X}} ). \label{eq:cov_hessian}
\eeq
Combining \eqref{eq:f_cov} and \eqref{eq:cov_hessian} yields
\beq
   S(\dot{\mbf{e}}) \tilde{X} = H_{f_I(\tilde{X})}(\dot{\tilde{X}}) - Q( S( \mbf{q}) , S(\mbf{e}) ) \tilde{X}. \label{eq:u_dot1}
\eeq
Resolving \eqref{eq:u_dot1} in basis $B_{\tilde{X}}$ yields
\beq
  \dot{\mbf{e}} = \mbf{H}(\tilde{X}) \mbf{q} - \mbs{\xi}, \nonumber
\eeq
where $S(\mbs{\xi}) = Q(S(\mbf{q}),S(\mbf{e}))$.
Recall that the Riemannian manifold under consideration, namely $(G,\langle \cdot, \cdot \rangle_X)$, is a smooth Riemannian manifold. This implies that the Levi-Civita connection is continuous \cite[p.115 Theorem 3.104]{book_Bullo_Lewis}. This further implies that the bilinear map $Q(\cdot,\cdot)$ is continuous and consequently $Q(\cdot,\cdot)$ is bounded, which is to say that there exists $m < \infty$ such that $ || Q(a,b) ||_\mathfrak{g} \le m ||a||_\mathfrak{g} || b ||_\mathfrak{g} $ for all $a,b \in \mathfrak{g}$. Thus, $|| S(\mbs{\xi}) ||_\mathfrak{g} \le m || S(\mbf{q}) ||_\mathfrak{g} || S(\mbf{e})||_\mathfrak{g}$, which implies that $ || \mbs{\xi} ||_2 \le m ||\mbf{q}||_2  || \mbf{e}||_2 $.
%
$\Box$

The proof of item (\ref{item:thm1_ic}) in Theorem \ref{thm:local_asymptotic_stability} continues as follows.
By Corollary 6.29 of \cite{book_Bullo_Lewis} there exists a constant $c > 0$ such that the set $\Omega_c = \{ \tilde{X} \in G \ | \ f_I(\tilde{X}) \le c \}$ is compact and the only critical point of $f$ in $\Omega_c$ is the point $\tilde{X} = I$. By Proposition 6.30 of \cite{book_Bullo_Lewis}, for all $c \in (0,L)$ there exists constants $0 < b_1 \le b_2$ such that
\beq
  b_1 ||\mbf{e}||_2^2 \le f(\hat{X},X) \le b_2 ||\mbf{e}||_2^2. \label{eq:quadratic}
\eeq

Consider again the function $V_1$ given in \eqref{eq:V1} and recall that $\dot{V}_1 \le 0$. Therefore, $V_1(\tilde{X}(t),\mbf{x}_f(t)) \le V_1(\tilde{X}(0),\mbf{x}_f(0))$ $\forall t \ge 0$. Let $V_1(\tilde{X}(0),\mbf{x}_f(0)) \le c$, where $c \in (0,L)$. Then, $f_I(\tilde{X}(t)) \le c$ $\forall t \ge 0$ and $\tilde{X} \in \Omega_c$ $\forall t \ge 0$. This implies that \eqref{eq:quadratic} is satisfied for all $t \ge 0$. Then, $V_1$ satisfies $\min\{ b_1 , \onehalf \underline{\lambda}(\mbf{P}_f) \} || \mbf{z} ||_2^2 \le V_1 \le \max \{ b_2 , \onehalf \bar{\lambda}(\mbf{P}_f) \} || \mbf{z} ||_2^2 $, where $\underline{\lambda}(\cdot)$ and $\bar{\lambda}(\cdot)$ respectively denote the maximum and minimum eigenvalues of a matrix, and $\mbf{z} = [ \ || \mbf{e} ||_2 \ ||\mbf{x}_f||_2 \ ]^\trans$. Thus, the fact that $V_1(\mbftilde{X}(t),\mbf{x}_f(t)) \le c$ for all $t \ge 0$ implies that $\mbf{e}$ and $\mbf{x}_f$ are bounded for all $t \ge 0$.

Consider the Lyapunov function candidate
\beq
  V_2(\tilde{X},\mbf{x}_f) = V_1(\tilde{X},\mbf{x}_f) - a \mbf{e}^\trans \mbf{C}_f \mbf{x}_f, \label{eq:V2}
\eeq
where $a \in (0,\infty)$.
Let $\gamma = || \mbf{C}_f ||_\frob$, then $V_2$ satisfies $\mbf{z}^\trans \mbf{W}_1 \mbf{z} \le V_2 \le \mbf{z}^\trans \mbf{W}_2 \mbf{z}$, where
\beq
  \mbf{W}_1 = \bma{c c} b_1 & -\onehalf a  \gamma \\ -\onehalf a  \gamma & \onehalf \underline{\lambda}(\mbf{P}_f) \ema, \ \text{and} \  \mbf{W}_2 = \bma{c c} b_2 & \onehalf a  \gamma \\ \onehalf a  \gamma & \onehalf \bar{\lambda}(\mbf{P}_f) \ema. \nonumber
\eeq
The derivative with respect to time of $V_2$ satisfies
\beqarray
  \dot{V}_2 & = & \dot{V}_1 - a \mbf{x}_f^\trans \mbf{C}_f^\trans \dot{\mbf{e}} - a \mbf{e}^\trans  \mbf{C}_f \dot{\mbf{x}}_f \nonumber \\
  & \le & - \onehalf \mbf{x}_f^\trans \mbf{Q}_f \mbf{x}_f - a \mbf{x}_f^\trans \mbf{C}_f^\trans  \dot{\mbf{e}} - a \mbf{e}^\trans  \mbf{C}_f \dot{\mbf{x}}_f. \nonumber
\eeqarray
From Corollary \ref{cor:u_f}, the derivative with respect to time of $\mbf{e}$ is given by $\dot{\mbf{e}} = -\mbf{H}(\tilde{X}) \mbf{u} - \mbs{\xi}$, where $S(\mbs{\xi}) = Q(S(\mbf{u}),S(\mbf{e}))$ and $|| \mbs{\xi} ||_2 \le m || \mbf{u} ||_2|| \mbf{e} ||_2$. Given that $\mbf{x}_f$ and $\mbf{e}$ are bounded, it follows that $\mbf{u}$ is bounded as well. As $\Omega_c$ is compact it follows that the norm of $\mbf{H}(\tilde{X})$ is bounded \cite{book_Bullo_Lewis}. Define $m_1 = \sup_{\tilde{X} \in \Omega_c} || \mbf{H}(\tilde{X}) ||_\frob$, $m_2 = m \cdot \sup|| \mbf{u} ||_2$, $\beta = ||\mbf{D}_f||_\frob$, and $\epsilon = \underline{\lambda}(\mbf{Q}_f)$. Then it can be shown that
\beqarray
  \dot{V_2} & \le & - \onehalf \epsilon ||\mbf{x}_f||^2 + a m_1 \gamma^2 ||\mbf{x}_f||_2^2 + a m_1 \beta \gamma ||\mbf{x}_f||_2 || \mbf{e}||_2 \nonumber \\
  &&{+} \: a \gamma m_2 ||\mbf{x}_f||_2  || \mbf{e}||_2 - a \mbf{e}^\trans \mbf{C}_f \mbf{A}_f \mbf{x}_f \nonumber \\
  &&{-} \: a\mbf{e}^\trans  \mbf{C}_f \mbf{B}_f  \mbf{e}. \nonumber
\eeqarray
Recall from \eqref{eq:kyp_C} $\mbf{C}_f = \mbf{B}_f^\trans \mbf{P}_f$. Let $\mbs{\Upsilon} =  \mbf{B}_f^\trans \mbf{P}_f \mbf{B}_f $ and note that since $\mbf{B}_f$ is assumed full rank and $\mbf{P}_f > 0$ it follows that $\mbs{\Upsilon} > 0$. Define $\delta =  ||  \mbf{A}_f||_\frob$, then
\begin{align}
  \dot{V}_2 & \le -(\onehalf \epsilon - a m_1 \gamma^2) ||\mbf{x}_f||_2^2  -  a \underline{\lambda}(\mbs{\Upsilon}) ||\mbf{e}||_2^2 \nonumber \\
  & \phantom{{}\le{}} + a \gamma (m_2 +  m_1 \beta + \delta)  ||\mbf{x}_f||_2 || \mbf{e} ||_2. \label{eq:V_2_dot}
\end{align}
Equation \eqref{eq:V_2_dot} is equivalent to $\dot{V}_2 \le - \mbf{z}^\trans \mbf{W}_3 \mbf{z}$, where
\beq
  \mbf{W}_3 = \bma{c c} a \underline{\lambda}(\mbs{\Upsilon})  & -  a \kappa  \\ -  a \kappa  & \onehalf \epsilon - a m_1 \gamma^2 \ema, \nonumber
\eeq
and $\kappa = \onehalf a \gamma (m_2 + m_1 \beta + \delta)$.
It can be shown that, provided
\beq
  a < \min \left \{  \sqrt{ \f{ 2 b_1 \underline{\lambda}(\mbf{P}_f) }{ \gamma^2 }  } , \f{ \onehalf \epsilon \underline{\lambda}(\mbs{\Upsilon})  }{ \underline{\lambda}(\mbs{\Upsilon}) m_1 \gamma^2 + \kappa^2 } \right \}, \nonumber
\eeq
the matrices $\mbf{W}_1$, $\mbf{W}_2$, and $\mbf{W}_3$ are positive definite. Therefore,
\beqarray
  \underline{\lambda}(\mbf{W}_1) ||\mbf{z}||_2^2 \le V_2 \le \bar{\lambda}(\mbf{W}_2) ||\mbf{z}||_2^2 \nonumber \\
  \dot{V}_2 \le - \underline{\lambda}(\mbf{W}_3) || \mbf{z}||_2^2, \nonumber
\eeqarray
which implies that
\beq
  ||\mbf{z}(t)||_2 \le \left ( \f{\underline{\lambda}(\mbf{W}_1)}{\bar{\lambda}(\mbf{W}_2)} \right)^{\onehalf} ||\mbf{z}(0)||_2 \exp \left( - \f{\underline{\lambda}(\mbf{W}_3)}{2 \bar{\lambda}(\mbf{W}_2)} t \right). \nonumber
\eeq
Therefore, trajectories of $(||\mbf{e}||_2,||\mbf{x}_f||_2)$ exponentially approach $(0,0)$. Due to the fact that trajectories of $\tilde{X}$ remain in $\Omega_c$ for all $t \ge 0$, $\mbf{e} \rightarrow \mbf{0}$ implies that $\tilde{X} \rightarrow I$ as $t \rightarrow \infty$. $\Box$

\section{Proof of Theorem \ref{thm:disturbance}} \label{app:theorem3}

The proof of item (\ref{item:thm2_as}) in Theorem \ref{thm:disturbance} follows in a similar manner to the proof of Theorem 5.1 in \cite{paper_khosravian_2015}. Consider the Lyapunov function candidate
\beq
  V_3(\tilde{X},\mbf{x}_f,\mbftilde{x}_d) = f(\tilde{X},\mbf{1}) + \mathcal{L}(\mbf{x}_f) + \onehalf \rho^{-1} \mbftilde{x}_d^\trans \mbftilde{x}_d. \label{eq:V3}
\eeq
The derivative with respect to time of the third term, $\onehalf \rho^{-1} \mbftilde{x}_d^\trans \mbftilde{x}_d$, is
\beq
  \f{\mathrm{d}}{\mathrm{d}t} \onehalf \rho^{-1} \mbftilde{x}_d^\trans \mbftilde{x}_d =  \onehalf \rho^{-1} \mbftilde{x}_d^\trans ( \mbf{A}_d^\trans + \mbf{A}_d ) \mbftilde{x}_d - \mbftilde{x}_d^\trans \mbf{C}_d^\trans  \mbfbar{e}. \nonumber
\eeq
Recall that $\mbf{A}_d$ is skew-symmetric and therefore $\mbf{A}_d^\trans  + \mbf{A}_d = \mbf{0}$. Consequently,
\beq
  \f{\mathrm{d}}{\mathrm{d}t} \onehalf \rho^{-1} \mbftilde{x}_d^\trans \mbftilde{x}_d  =   - \mbftilde{x}_d^\trans \mbf{C}_d^\trans \mbfbar{e} = -\mbftilde{w}^\trans  \mbfbar{e}. \nonumber
\eeq
The derivative with respect to time of $V_3$ can therefore be written as
\beq
  \dot{V}_3  =  \langle {\nabla}_{\tilde{X}} f(\tilde{X},I), \dot{\tilde{X}} \rangle_{\tilde{X}} + \dot{\mathcal{L}}(\mbf{x}_f) - \mbftilde{w}^\trans  \mbfbar{e}. \label{eq:V3_dot}
\eeq
Substituting \eqref{eq:Xtilde_dot} into \eqref{eq:V3_dot} yields
\beqarraynn
  \dot{V}_3 & = & \langle {\nabla}_{\tilde{X}} f(\tilde{X},I), \dot{\tilde{X}} \rangle_{\tilde{X}} + \dot{\mathcal{L}}(\mbf{x}_f) - \mbftilde{w}^\trans  \mbfbar{e} \\
  & = & \langle S(\mbf{e}) \tilde{X}, \tilde{X} \Ad_X(\tilde{w})  \rangle_{\tilde{X}} - \langle {S}(\mbf{e}) \tilde{X}, {S}(\mbf{u}) \tilde{X} \rangle_{\tilde{X}}  \\
  &&{+} \:  \dot{\mathcal{L}}(\mbf{x}_f) - \mbftilde{w}^\trans  \mbfbar{e} \\
  & = &  - \langle {S}(\mbf{e}), {S}(\mbf{u}) \rangle + \langle {S}(\mbf{e}), \Ad_{\hat{X}}(\tilde{w}) \rangle \\
  &&{+} \:  \dot{\mathcal{L}}(\mbf{x}_f) - \mbftilde{w}^\trans \mbfbar{e} \\
  & = & -\mbf{e}^\trans  \mbf{u} + \langle \Ad_{\hat{X}}^*({S}(\mbf{e})),{S}(\mbftilde{w})  \rangle + \dot{\mathcal{L}}(\mbf{x}_f) - \mbftilde{w}^\trans  \mbfbar{e}\\
  & = & -\mbf{e}^\trans  \mbf{u} + \langle S(\mbfbar{e}) , S(\mbftilde{w})  \rangle + \dot{\mathcal{L}}(\mbf{x}_f) - \mbftilde{w}^\trans  \mbfbar{e} \\
  & = & -\mbf{e}^\trans  \mbf{u} + \mbfbar{e}^\trans  \mbftilde{w} + \dot{\mathcal{L}}(\mbf{x}_f) - \mbftilde{w}^\trans  \mbfbar{e} \\
  & = &  -\mbf{e}^\trans  \mbf{u}  + \dot{\mathcal{L}}(\mbf{x}_f) .
\eeqarraynn
By Corollary \ref{cor:spr}, $\dot{V}_3 = -\onehalf \mbf{x}_f^\trans \mbf{Q}_f \mbf{x}_f - \mbf{e}^\trans  \mbf{D}_f  \mbf{e}$.
Therefore, $\dot{V}_3 \le 0$. As $f$ is an error function it follows that $f$ is locally positive definite and $\tilde{X} = I$ is an isolated critical point of $f$ \cite{book_Bullo_Lewis}.
By assumption, there exists a faithful representation of $G$ as a matrix Lie group. This implies that there exists $m>0$ and a mapping $\mbs{\Phi} : G \rightarrow GL(m)$ such that $\mbs{\Phi}(G)$ is a matrix Lie group \cite{paper_khosravian_2015}.
Following the proof of Theorem 5.1 in \cite{paper_khosravian_2015}, this implies that there exists a set $B_r = \{ \tilde{X} \in G \ | \ d(\tilde{X}) \le r \} $ about $\tilde{X} = I$, where $d(\tilde{X}) = || \mbf{1} - \mbs{\Phi}(\tilde{X})||_\frob$, such that for all $\tilde{X} \in B_r$, $f(\tilde{X},I)$ is positive definite and $\tilde{X}= I$ is the only critical point of $f$ in $B_r$. Further, this implies that $V_3(\tilde{X},\mbf{x}_f,\mbftilde{x})$ is positive definite in the set $\bar{B}_r = \{ (\tilde{X},\mbf{x}_f,\mbftilde{x}_d) \in G \times \mathbb{R}^{n_f} \times \mathbb{R}^{n_d} \ | \ \ell(\tilde{X},\mbf{x}_f,\mbftilde{x}_d) \le r \}$, where $\ell(\tilde{X},\mbf{x}_f,\mbftilde{x}_d) = d(\tilde{X}) + ||\mbf{x}_f||_2 + ||\mbftilde{x}_d||_2$, and $\tilde{X} = I$ is the only critical point of $f$ in the set $\bar{B}_r$. Consequently, for all $t \ge 0$ and all $(\tilde{X},\mbf{x}_f,\mbftilde{x}_d) \in \bar{B}_r$ the Lyapunov function $V_3$ is positive definite and $\dot{V}_3 \le 0$ and by Theorem 4.8 of \cite{khalil2002nonlinear} the equilibrium point $(\tilde{X},\mbf{x}_f,\mbftilde{x}_d) = (I,\mbf{0},\mbf{0})$ is uniformly stable.

Again, following the proof of Theorem 5.1 in \cite{paper_khosravian_2015}, choose $\alpha < \min_{\ell(\tilde{X},\mbf{x}_f,\mbftilde{x}_d) = r} V_3(\tilde{X},\mbf{x}_f,\mbftilde{x}_d)$, and define $\mathscr{L}_\alpha = \{ (\tilde{X},\mbf{x}_f,\mbftilde{x}_d) \in \bar{B}_r \ | \ V_3(\tilde{X},\mbf{x}_f,\mbftilde{x}_d) \le \alpha \}$. Then, $\mathscr{L}_\alpha \subset \bar{B}_r$ and all trajectories of $(\tilde{X},\mbf{x}_f,\mbftilde{x}_d)$ starting in $\mathscr{L}_\alpha$ remain in $\mathscr{L}_\alpha$ for all $t \ge 0$ \cite{khalil2002nonlinear}. Consequently, trajectories of $(\tilde{X},\mbf{x}_f,\mbftilde{x}_d)$ remain bounded with respect to $\ell(\tilde{X},\mbf{x}_f,\mbftilde{x}_d)$, which implies $\tilde{X}$ remains bounded with respect to $d(\cdot)$ and $\mbf{x}_f,\mbftilde{x}_d$ remain bounded with respect to $||\cdot||_2$.


As $(\tilde{X},\mbf{x}_f,\mbftilde{x}_d) \in \mathscr{L}_\alpha$ for all $t \ge 0$, it follows that $\tilde{X} \in \Omega_\alpha$ for all $t \ge 0$, where $\Omega_\alpha = \{ {X} \in B_r \ | \ f(\tilde{X},I) \le \alpha \}$. Moreover, $\mathscr{L}_\alpha \subset \bar{B}_r$ implies that the only critical point of $f$ in $\mathscr{L}_\alpha$ is $\tilde{X} = I$ and therefore the only critical point in $\Omega_\alpha$ is $\tilde{X} = I$. Consequently, $\Omega_\alpha \subset \Omega_L$ and therefore by Proposition 6.30 of \cite{book_Bullo_Lewis} there exists constants $b_1$ and $b_2$ such that \eqref{eq:quadratic} is satisfied. This implies that $\mbf{e}$ is bounded for all $t \ge 0$.

%



Taking the second derivative with respect to time of $V_3$ yields $\ddot{V}_3 = - \mbf{x}_f^\trans \mbf{Q}_f \dot{\mbf{x}}_f - 2 \mbf{e}^\trans \mbf{D}_f  \dot{\mbf{e}}$. By Corollary \ref{cor:u_f}, $\dot{\mbf{e}} = -\mbf{H}(\tilde{X}) \mbf{q} - \mbs{\xi}$ where $S(\mbf{q}) = \Ad_{\hat{X}}({S}(\mbftilde{w})) + S(\mbf{u})$ and ${S}(\mbs{\xi}) = {Q}({S}(\mbf{q}),S(\mbf{e}))$. Given that $\tilde{X}$, $\mbf{x}_f$, $\mbftilde{x}_d$, and $\mbf{e}$ are bounded and ${X}$ is bounded by assumption, it follows that $\dot{\mbf{x}}_f$ and $\dot{\mbf{e}}$ are bounded. Thus, $\ddot{V}_3$ is bounded and therefore $\dot{V}_3$ is uniformly continuous. By application of Barbalat's Lemma $\dot{V}_3 \rightarrow 0$ as $t \rightarrow \infty$ and thus $\mbf{x}_f \rightarrow \mbf{0}$ as $t \rightarrow \infty$. Taking the derivative with respect to time of $\dot{\mbf{x}}_f$ gives $\ddot{\mbf{x}}_f = \mbf{A}_f^2 \mbf{x}_f + \mbf{A}_f \mbf{B}_f  \mbf{e} + \mbf{B}_f  \dot{\mbf{e}}$, which is bounded. Applying Barbalat's Lemma, $\dot{\mbf{x}}_f \rightarrow \mbf{0}$. This fact, along with the assumption that $\mbf{B}_f$ has full rank implies that $\mbf{e} \rightarrow \mbf{0}$ and ${\nabla}_{\tilde{X}} f(\tilde{X},I) \rightarrow 0$. Consequently, $\mbf{u} \rightarrow \mbf{0}$ as $t \rightarrow \infty$. Since the only critical point of $f$ in $\mathscr{L}_{\alpha}$ is the point $\tilde{X} = I$, it follows that $\tilde{X} \rightarrow I$ as $t \rightarrow \infty$. Due to the boundedness of $\tilde{X}$, $\mbf{x}_f$, and $\mbftilde{x}_d$ as well as the assumption that ${X}$ and $v$ are bounded it can be shown that $\ddot{\tilde{X}}$ is bounded. Thus $\dot{\tilde{X}}$ is uniformly continuous and by Barbalat's Lemma $\dot{\tilde{X}} \rightarrow 0$ as $t \rightarrow \infty$. From \eqref{eq:error_disturbance},
\beqarraynn
  0 & = & \lim_{t\rightarrow \infty} \left ( \tilde{X} \Ad_X({S}(\mbftilde{w}))  - \mbf{S}(\mbf{u}) \tilde{X} \right) \\
  & = & \lim_{t\rightarrow \infty}  \Ad_X({S}(\mbftilde{w})) 
\eeqarraynn
and thus $\mbftilde{w} \rightarrow \mbf{0}$ as $t \rightarrow \infty$. With the assumption that $\mbf{C}_d$ has full rank $\mbftilde{w} \rightarrow \mbf{0}$ implies that $\mbftilde{x}_d \rightarrow \mbf{0}$ as $t \rightarrow \infty$.

It has been shown that for all $(\tilde{X}(0),\mbf{x}_f(0),\mbftilde{x}_d(0)) \in \mathscr{L}_\alpha$, $(\tilde{X},\mbf{x}_f,\mbftilde{x}_d) \rightarrow (\mbf{1},\mbf{0},\mbf{0})$ as $t \rightarrow \infty$.
As $V_3$ is positive definite on $\mathscr{L}_\alpha$ there exists a class $\mathcal{K}$ function $\phi$ such that $\phi(\ell(\tilde{X},\mbf{x}_f,\mbftilde{x}_d)) \le V(\tilde{X},\mbf{x}_f,\mbftilde{x}_d)$ for all $(\tilde{X},\mbf{x}_f,\mbftilde{x}_d) \in \mathscr{L}_\alpha $ \cite{khalil2002nonlinear}. Thus, $(\tilde{X},\mbf{x}_f,\mbftilde{x}_d) \rightarrow (I,\mbf{0},\mbf{0})$ as $t \rightarrow \infty$ for all $ \ell(\tilde{X}(0),\mbf{x}_f(0),\mbftilde{x}_d(0)) < \phi^{-1}(\alpha) $ and therefore the equilibrium point $(\tilde{X},\mbf{x}_f,\mbftilde{x}_d) = (I,\mbf{0},\mbf{0})$ is uniformly convergent \cite{book_marquez}. This, along with the fact that the equilibrium point is uniformly stable, shows that $(\tilde{X},\mbf{x}_f,\mbftilde{x}_d)=(I,\mbf{0},\mbf{0})$ is locally uniformly asymptotically stable.

To show item (\ref{item:thm2_ic}) of Theorem \ref{thm:disturbance}, recall that $\dot{V}_3 \le 0$ and consequently $V_3(\tilde{X}(t),\mbf{x}_f(t),\mbftilde{x}_d(t)) \le V_3(\tilde{X}(0),\mbf{x}_f(0),\mbftilde{x}_d(0))$ for all $t \ge 0$. By assumption $V_3(\tilde{X}(0),\mbf{x}_f(0),\mbftilde{x}_d(0)) < L$, and therefore $V_3(\tilde{X}(t),\mbf{x}_f(t),\mbftilde{x}_d(t)) < L$ for all $t \ge 0$. This implies that $\mbf{x}_f$ and $\mbftilde{x}_f$ remain bounded for all $t \ge 0$ and that \eqref{eq:quadratic} is satisfied for all $t \ge 0$, which implies that $\mbf{e}$ is bounded for all $t \ge 0$. Applying Barbalat's Lemma on $V_3$, $\mbf{x}_f$, and $\tilde{X}$, as above, it can be shown that $\mbf{e} \rightarrow \mbf{0}$, $\mbf{x}_f \rightarrow \mbf{0}$, and $\mbftilde{x}_d \rightarrow \mbf{0}$ as $t \rightarrow \infty$. Due to the fact that $V_3(\tilde{X}(t),\mbf{x}_f(t),\mbftilde{x}_d(t)) < L$, $\mbf{e} \rightarrow \mbf{0}$ implies that $\tilde{X} \rightarrow I$. Therefore trajectories of $(\tilde{X},\mbf{x}_f,\mbftilde{x}_d)$ asymptotically approach $(I,\mbf{0},\mbf{0})$. $\Box$
\section*{Acknowledgment}

The authors would like to thank the reviewers whose input has greatly improved the paper.


\addcontentsline{toc}{section}{References}
\bibliographystyle{ieeetr}
\bibliography{ref_forbes}

\begin{IEEEbiography}[{\includegraphics[width=1in,height=1.25in,clip,keepaspectratio]{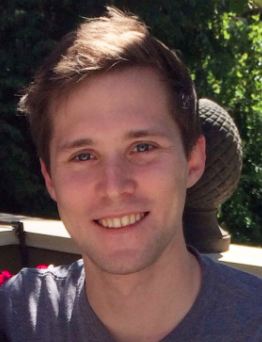}}]{David Evan Zlotnik}
received his Bachelors of Mechanical Engineering (Honours) from McGill University in 2013 and his MSE in Aerospace Engineering at the University of Michigan in 2016.
David is currently a Ph.D.\ Candidate in the department of Aerospace Engineering at the University of Michigan.
His research interests include nonlinear observer design, control of flexible space structures and robotic manipulators, as well as localization and mapping for mobile robotics.
\end{IEEEbiography}
\vfill

\vspace{-50mm}

\begin{IEEEbiography}[{\includegraphics[width=1in,height=1.25in,clip,keepaspectratio]{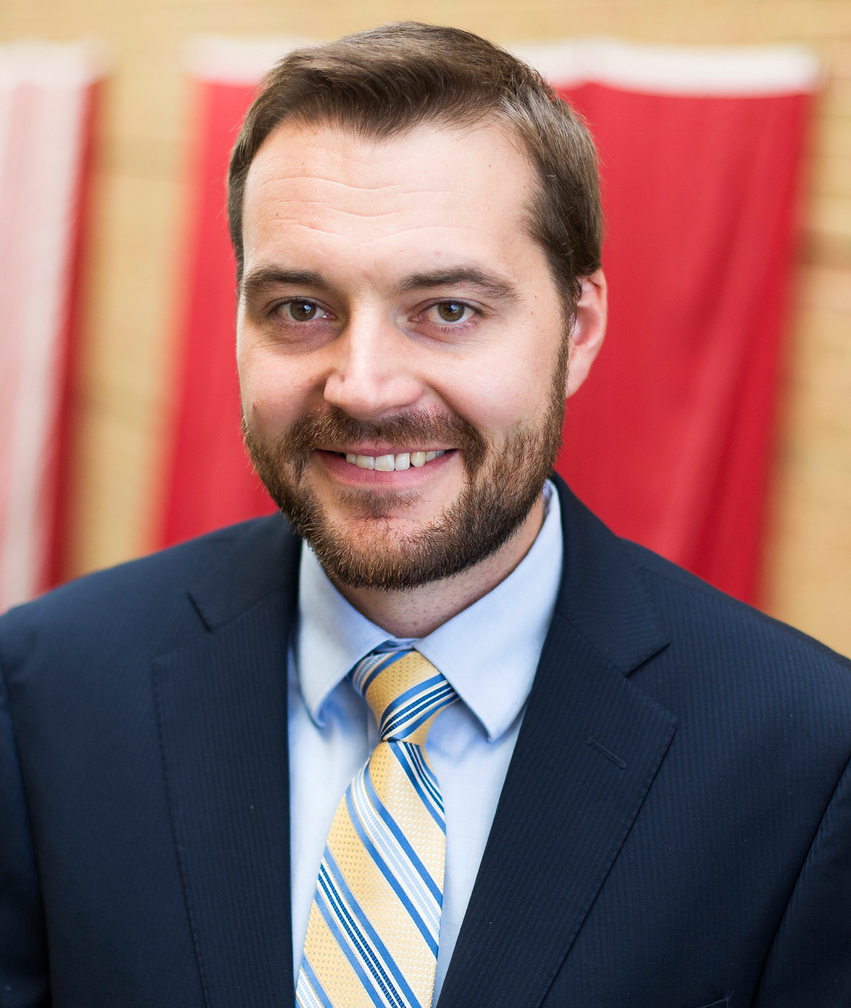}}]{James Richard Forbes}
received his B.A.Sc. in Mechanical Engineering
(Honours, Co-op) from the University of Waterloo in 2006. While
attending the University of Waterloo James participated in the co-op
program; James had the opportunity to work in the manufacturing,
automotive, rail, and industrial automation (robotics) industries.
James was awarded his M.A.Sc. and Ph.D. degrees in Aerospace Science
and Engineering from the University of Toronto Institute for Aerospace
Studies (UTIAS) in 2008 and 2011, respectively. He was awarded the G.
N. Patterson Award for the most outstanding Ph.D. thesis in 2011. With
Anton de Ruiter and Christopher Damaren, James coauthored the text
``Spacecraft Dynamics and Control -- An Introduction'' published by Wiley
in 2013 (ISBN-13: 978-1118342367). From May 2011 to August 2013 James
was an Assistant Professor of Mechanical Engineering at McGill
University located in Montreal, Quebec, Canada. From September 2013 to
July 2015 James was an Assistant Professor of Aerospace Engineering at
the University of Michigan. In August 2015 James returned to McGill
University as an Assistant Professor of Mechanical Engineering. James
is also a member of McGill's Center for Intelligent Machines (CIM).
James' research interests include nonlinear, robust, and optimal
estimation and control as applied to robotic and aerospace systems.
\end{IEEEbiography}

\vfill




\end{document}